\documentclass[a4paper,11pt]{article}

\usepackage{jheppub}
\usepackage{graphicx}
\usepackage{amsmath}
\usepackage{slashed}
\usepackage{braket}
\usepackage[export]{adjustbox}
\usepackage{enumitem}
\usepackage{placeins}
\usepackage[normalem]{ulem}
\usepackage{xspace}
\usepackage{caption}
\usepackage{subcaption}

\newcommand{\ord}{\begin{cal}O\end{cal}}
\def\bsp#1\esp{\begin{split}#1\end{split}}
\newcommand{\df}{\mathrm{d}}

\newcommand{\eps}{\epsilon}

\newcommand{\cM}{\mathcal{M}}

\def\beq{\begin{equation}}
\def\eeq{\end{equation}}
\def\bea{\begin{eqnarray}}
\def\eea{\end{eqnarray}}

\title{Lepton-pair production at hadron colliders at N$^3$LO in QCD}

\author[a]{Claude Duhr}
\emailAdd{cduhr@uni-bonn.de}

\author[b]{Bernhard Mistlberger}
\emailAdd{bernhard.mistlberger@gmail.com}

\affiliation[a]{Bethe Center for Theoretical Physics, Universit\"at Bonn, D-53115, Germany}
\affiliation[b]{SLAC National Accelerator Laboratory, Stanford University, Stanford, CA 94039, USA}

\abstract{
We compute for the first time the complete corrections at N$^3$LO in the strong coupling constant to the inclusive neutral-current Drell-Yan process including contributions from both photon and $Z$-boson exchange. 
Our main result is the computation of the QCD corrections to the inclusive production cross section of an axial-vector boson to third order in the strong coupling in a variant of QCD with five massless quark flavours. 
Since the axial anomaly does not cancel for an odd number of flavours, we also consistently include non-decoupling effects in the top-quark mass through three loops. 
We perform a phenomenological study of our results, and we present for the first time predictions for the inclusive Drell-Yan process at the LHC at this order in QCD perturbation theory.
}

\keywords{N$^3$LO, Drell-Yan production.}
\preprint{\begin{minipage}[t]{8cm}\begin{flushright}BONN-TH-2021-12, SLAC-PUB-17632\\
          \end{flushright}\end{minipage}}

\begin{document}

\maketitle

\section{Introduction}
The production of a pair of massless leptons at hadron colliders like the Large Hadron Collider (LHC) at CERN -- the so-called Neutral-Current Drell-Yan (NCDY) process -- is one of the most important and most studied hadron collider processes. From an experimental perspective, its clean final-state signature makes it an
ideal candidate for luminosity measurements and detector calibration. The DY process also plays a key
role in the measurement of parton distribution functions
(PDFs) at the LHC and in many searches for physics
beyond the Standard Model (SM). From a theoretical perspective, the invariant-mass distribution of the produced lepton pair is arguably the simplest hadron collider observable, and it often serves as a template to understand the structure of higher-order QCD corrections at hadron colliders more generally. It is thus important to have a solid theoretical understanding for the NCDY process, including higher-order corrections in both the strong and electroweak coupling constants.

At leading order (LO) in the electroweak coupling constant, the massless lepton pair is produced via the propagation of an intermediate, off-shell photon or $Z$-boson~\cite{Drell:1970wh}. 
The next-to-leading order (NLO) QCD corrections have been computed several decades ago~\cite{Altarelli:1978id,Altarelli:1979ub}. 
The next-to-next-to-leading order (NNLO) corrections were computed in refs.~\cite{Matsuura:1987wt,Matsuura:1988nd,Matsuura:1988sm,Hamberg:1990np,Matsuura:1990ba,vanNeerven:1991gh}  
in a version of the SM where all quarks, including the top quark, are considered massless, and the effects of finite quark masses were computed in refs.~\cite{Dicus:1985wx,Gonsalves:1991qn,Rijken:1995gi}. Fully differential predictions for the Drell-Yan process at NNLO are readily available, see for example refs.~\cite{Anastasiou:2003yy,Cieri:2015rqa,Gavin:2010az,Boughezal:2016wmq,Grazzini:2017mhc}.
Electroweak corrections, including mixed QCD-electroweak corrections are also available~\cite{Delto:2019ewv,Bonciani:2019nuy,Bonciani:2020tvf,Bonciani:2021zzf,Buccioni:2020cfi}.

The uncertainty on the NNLO results due to missing higher orders in the QCD perturbative expansion was estimated to be at the percent level by varying the renormalisation and factorisation scales by a factor of two around a central scale related to the invariant mass of the lepton pair.
Very recently, also next-to-next-to-next-to-leading order (N$^3$LO) corrections in the strong coupling have been computed, but only for the contributions from an intermediate off-shell-photon~\cite{Duhr:2020seh}. Unlike for N$^3$LO corrections to inclusive Higgs production processes ~\cite{Anastasiou:2015ema,Anastasiou:2016cez,Duhr:2019kwi,Chen:2019lzz,Duhr:2020kzd,Dreyer:2018qbw,Dreyer:2016oyx}, it was found that for large ranges of invariant masses, the N$^3$LO corrections are sizeable and lower the value of the cross section by a few percent, which is more than the missing higher order uncertainty estimated from varying the perturbative scales at NNLO. This result was recently confirmed by the independent computation from the double-differential computation for lepton-pair production via an off-shell photon at N$^3$LO in ref.~\cite{Chen:2021vtu} and the computation of the fiducial cross setion at N$^3$LO in ref.~\cite{Camarda:2021ict}. A similar behaviour was observed for the production of a lepton-neutrino pair at N$^3$LO~\cite{Duhr:2020sdp}. This shows that N$^3$LO corrections are highly needed if we want to reach a precision at the percent level for vector-boson production at hadron colliders. In particular, a complete calculation of the NCDY process in the SM at N$^3$LO, including also the contributions from the exchange of the $Z$-boson, is highly desired.

The computation of higher-order corrections including $Z$-bosons, however, is much more challenging. Unlike the photon, the $Z$-boson has both vector and axial couplings to fermions. Higher-order computations typically diverge, and they are conventionally regularised using dimensional regularisation, where the space-time dimension is analytically continued from $D=4$ to $D=4-2\eps$ dimensions. Axial couplings involve a $\gamma^5$ matrix, which is an intrinsically four-dimensional object, and so its treatment in dimensional regularisation is ambiguous (see ref.~\cite{Gnendiger:2017pys} and references therein for a review). Moreover, it is well known that the axial current is anomalous in QCD, and its divergence is described by the famous Adler-Bell-Jackiw (ABJ) anomaly equation~\cite{Adler:1969gk,Bell:1969ts,Adler:1969er} (this anomaly is absent for the vector current, as a consequence of Yang's theorem~\cite{Yang:1950rg}). In the complete SM, the axial anomaly cancels. Calculations involving massive top quarks, however, are technically extremely challenging, and therefore computations are usually done in an effective theory where the top quark is infinitely heavy and is integrated out. This procedure does no longer naively work in the presence of an axial current, because the resulting effective theory is anomalous~\cite{Collins:1978wz}.

The goal of this paper is to present for the first time complete results for the invariant mass distribution for the NCDY process in the SM at N$^3$LO in the strong coupling constant, including contributions from both an intermediate photon and $Z$-boson. Our main contribution is the computation the inclusive N$^3$LO cross section for an axial-vector state in QCD. We treat the $\gamma^5$ matrix by working in the Larin scheme~\cite{tHooft:1972tcz,Breitenlohner:1977hr,Larin:1991tj,Larin:1993tq,Jegerlehner:2000dz} and we work in a theory with five massless active quark flavours where the top quark has been integrated out. Since this effective theory is anomalous, the top quark does not completely decouple. We include non-decoupling effects via a Wilson coefficient multiplying the axial current in the effective theory. As a result, we obtain the complete N$^3$LO cross section in the SM, up to terms that are power-suppressed in the top quark mass. The finite top-mass effects are known to be small at NNLO, and we expect the size of the missing power-suppressed terms to be negligible at N$^3$LO as well. As a result, we obtain for the first time complete phenomenological predictions for the NCDY process to third order in the strong coupling.

This paper is organised as follows: In section~\ref{sec:setup} we review the structure of the QCD corrections to the NCDY process. In section~\ref{sec:axial} we present the main ingredient of our computation, namely the cross section for axial-vector production at N$^3$LO in QCD, and we discuss our treatment of the $\gamma^5$ matrix in dimensional regularisation and the non-decoupling top-mass contributions.
In section~\ref{sec:pheno} we discuss the phenomenological implications of our results. In section~\ref{sec:conclusions} we draw our conclusions.


\section{The Neutral-Current Drell-Yan process}
\label{sec:setup}


The Neutral-Current Drell-Yan (NCDY) process describes the production of a pair of (massless) leptons 
as the result of the collision of (anti-) protons:
\beq\label{eq:NCDY_process}
\text{P}(P_1)\,+\,\text{P}(P_2) \to l \, \bar l(Q) +X\,.
\eeq
Here, $P_1$ and $P_2$ are the momenta of the scattering protons, $Q$ is the invariant mass of the lepton pair $l\,\bar{l}$, and $X$ collectively denotes additional QCD radiation.
The goal of this paper is to compute this process to third order in the strong coupling constant and to discuss the associated phenomenology. More precisely, we want to compute the cross section for the process in eq.~\eqref{eq:NCDY_process} differentially in the invariant mass $Q$. The production cross section is mediated by a virtual photon or $Z$-boson which subsequently decay to the final-state leptons. Our formalism is accurate up to corrections that are suppressed in the electroweak coupling constant $\alpha_{EW}$:
\beq
Q^2 \frac{\df \sigma_{P\, P\rightarrow l \bar l+X}}{\df Q^2}=\sum\limits_{B,B^\prime \in \{\gamma^*,Z\}} Q^2 \frac{\df \sigma_{P\, P\to B / B^\prime +X \to l \bar l+X}}{\df Q^2}+\mathcal{O}\left(\alpha_{EW}^3\right).
\eeq

 \begin{figure}[!h]
  \begin{center}
  \includegraphics[scale=0.15]{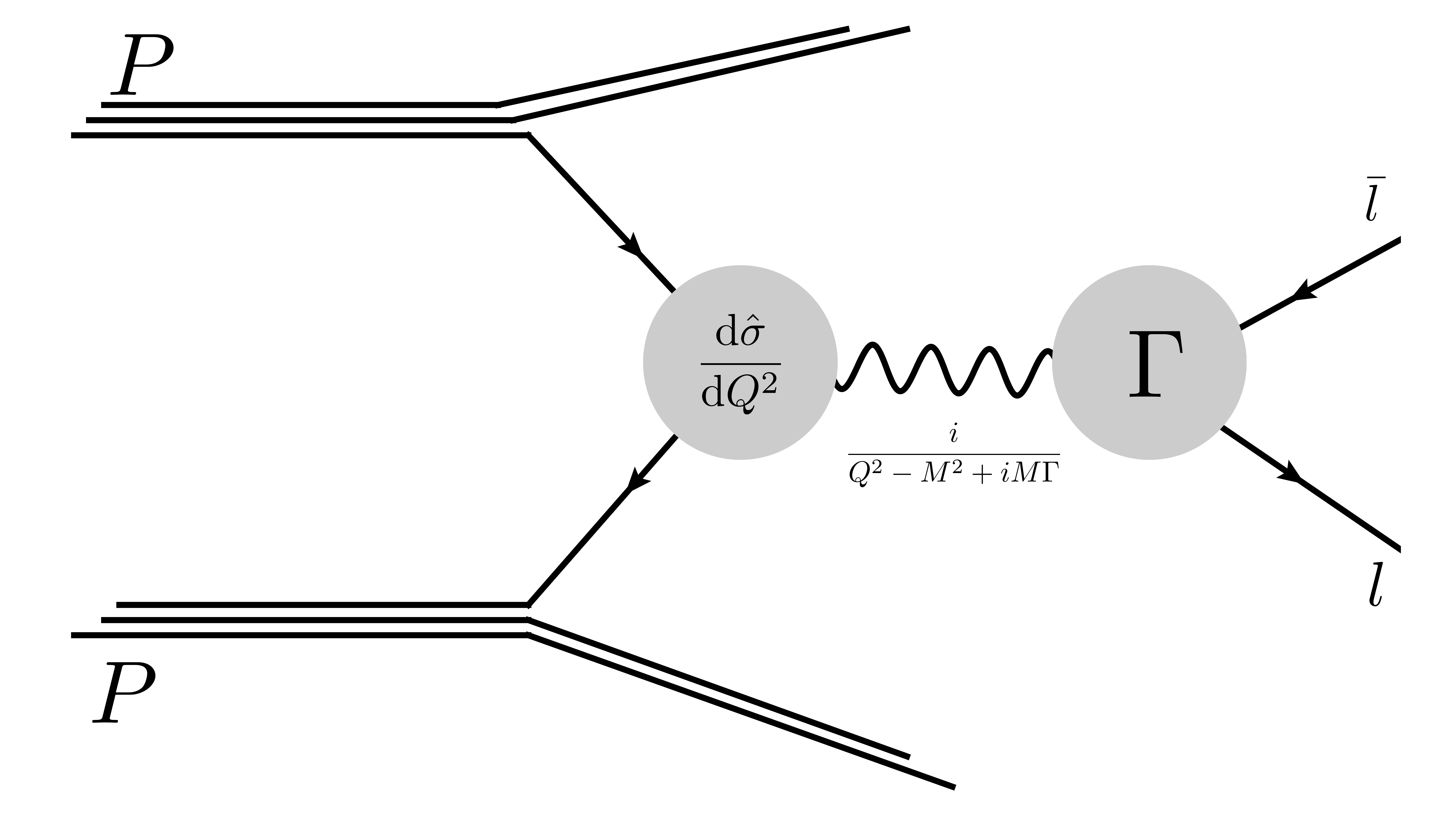}
  \caption{\label{fig:masterform}
Schematic depiction of the Drell-Yan process and its factorisation into the production probability of a virtual gauge boson and the subsequent decay to final-state leptons.
  }
  \end{center}
  \end{figure} 

Figure~\ref{fig:masterform} schematically shows the NCDY process for the production of a lepton pair via an intermediate vector boson. The cross section can be written in a factorised form,
\beq
\label{eq:master}
Q^2 \frac{\df \sigma_{P\, P\to B / B^\prime +X \to l\,\bar{l}+X}}{\df Q^2}= Q^2 \frac{\df \sigma_{P\, P\rightarrow B / B^\prime+X}}{\df Q^2} \times \Gamma_{B / B^\prime \to l\,\bar{l}}(Q^2) \times \text{BW}(Q^2,m_B,m_{B^\prime})\,.
\eeq
We now describe the factors in the right-hand side of eq.~\eqref{eq:master} in detail. 

The propagation of the virtual gauge bosons is described by the Breit-Wigner distribution,
\beq
\text{BW}(Q^2,m_B,m_B^\prime) =\frac{Q^3}{\pi} \mathtt{Re}\left[\frac{1}{(Q^2-m_{B}^2)+i m_B \Gamma_B }\frac{1}{(Q^2-m_{{B^\prime}}^2)-i m_{B^\prime} \Gamma_{B^\prime} }\right]\,.
\eeq
The decay of the gauge bosons into leptons is captured by the interference contributions to the width:
\bea\label{eq:decay}
&&\Gamma_{B / B^\prime \to l\bar l}(Q ,m_l, m_{\bar l}) =\\
&&\,\, \frac{1}{6Q}\int \frac{\df ^d p_{l}}{(2\pi)^{d-1}}\delta_+(p_l^2-m_l^2)\frac{\df ^d p_{\bar l}}{(2\pi)^{d-1}}\delta_+(p_{\bar l}^2-m_{\bar l}^2) (2\pi)^d \delta^d(q-p_l-p_{\bar l})
 \mathcal{M}_{B \to l\bar l} \cdot \mathcal{M}_{B^\prime \to l\bar l}^*\,, \nonumber
\eea
We use the notation $\cM_1\cdot \cM_2^*$ to indicate that the interferences are summed over the colours and spins of all external particles. 
Equation~\eqref{eq:decay} describes the interferences of the decay amplitudes of gauge bosons into two final-state particles with masses $m_l$ and $m_{\bar l}$ at Born level. It is chosen such that if the formula is evaluated on-shell, $Q=m_B$ and $B=B^\prime$, then it corresponds to the partial width $\Gamma_{B\to l\bar l}$ of the gauge boson $B$ in its rest-frame. 
Since we are working to leading order in the electroweak coupling, we only need to consider the tree-level contributions to the decay. For massless leptons, we find:
\beq\label{eq:massless_decay}
\Gamma_{B / B^\prime \to l \bar l}(Q^2) := \Gamma_{B / B^\prime \to l \bar l}(Q^2,0,0)=\frac{ Q m_W^2}{ 3\pi v^2}\left[Q_{V,l}^{B}Q_{V,l}^{B^\prime}+Q_{A,l}^{B}Q_{A,l}^{B^\prime}\right]\, .
\eeq
The charges in the above equation for a $Z$ boson or photon are defined as :
\bea
Q_{V,f}^{\gamma^*}&=&Q_f \sqrt{1-\frac{m_W^2}{m_Z^2}}\,,\hspace{1cm} 
Q_{A,f}^{\gamma^*}=0\,,\\
Q_{V,f}^{Z}&=&\frac{m_Z}{2m_W} \left[T_f^3-2 Q_f\left(1-\frac{m_W^2}{m_Z^2}\right)\right]\,,\hspace{1cm}
Q_{A,f}^{Z}=\frac{m_Z}{2m_W} T_f^3\,.\nonumber
\eea
Here, $m_W$ and $m_Z$ are the masses of the $W$ and $Z$ bosons.
In the SM we have the following charge assignments for up-type and down-type quarks and electrically-charged leptons:
\beq
\begin{array}{c|c|c|c}
\hline\hline
 & \hspace{0.5cm}u \hspace{0.5cm} &\hspace{0.5cm} d \hspace{0.5cm} & \hspace{0.5cm} l   \hspace{0.5cm}\\
 \hline
 Q_f & \frac{2}{3} & -\frac{1}{3}  & -1  \\
 T_f^3 & \frac{1}{2} & -\frac{1}{2} & -\frac{1}{2} \\
 \hline\hline
\end{array}
\eeq

Finally, the first factor in eq.~\eqref{eq:master} describes the production cross section for the interference of two off-shell gauge bosons $B$ and $B^\prime$:
\beq\bsp
Q^2 \frac{\df \sigma_{P\, P\rightarrow B / B^\prime+X}}{\df Q^2}=\tau \int_0^1 \df x_1\, \df x_2\, \df z\, \delta(\tau-x_1 x_2 z) \, \sum_{i,j} f_i(x_1)\,f_j(x_2)\,  \eta_{\, ij\to B / B^\prime +X}(z)\,.
\esp\eeq
In the previous equation we introduced the parton distribution functions $f_i(x)$ that are convoluted with the partonic coefficient functions $ \eta_{\, ij\to B / B^\prime +X}$. We suppress the dependence of all quantities on the factorisation scale to improve the readability.
The variables $x_{1,2}$ represent the fraction of the momenta of the protons carried by the initial-state partons.
We introduce the variables 
\beq
S=(P_1+P_2)^2,\hspace{1cm}\tau=\frac{Q^2}{S}.
\eeq
The partonic coefficient functions are given by
\beq
\eta_{\, ij\to B / B^\prime +X} =\frac{\mathcal{N}_{ij}}{2Q^2}  \sum_{X_i}\int \df \Phi_{B / B^\prime +X_i} \mathcal{M}_{ij\to B  +X_i} \cdot \mathcal{M}_{ij\to B^\prime +X_i}^*\,,
\eeq
where the factor $\mathcal{N}_{ij}$ represents a process-dependent averaging over initial-state spins and colours,  $ \df \Phi_{B / B^\prime +X_i}$ is the phase space measure for a particular final state, and $\mathcal{M}_{ij\to B+X_i}$ is the scattering matrix element for the production of this final state.  
In order to compute the partonic coefficient functions to a given order in perturbation theory, we consider a perturbative expansion of the product of matrix elements truncated at a fixed order in the coupling constant.
The partonic coefficient functions are the main ingredients to our computation, as they incorporate the entirety of the higher-order QCD corrections. We will study their structure in more detail in the remainder of this section.

\begin{figure}[!h]
  \begin{center}
  \includegraphics[scale=0.1]{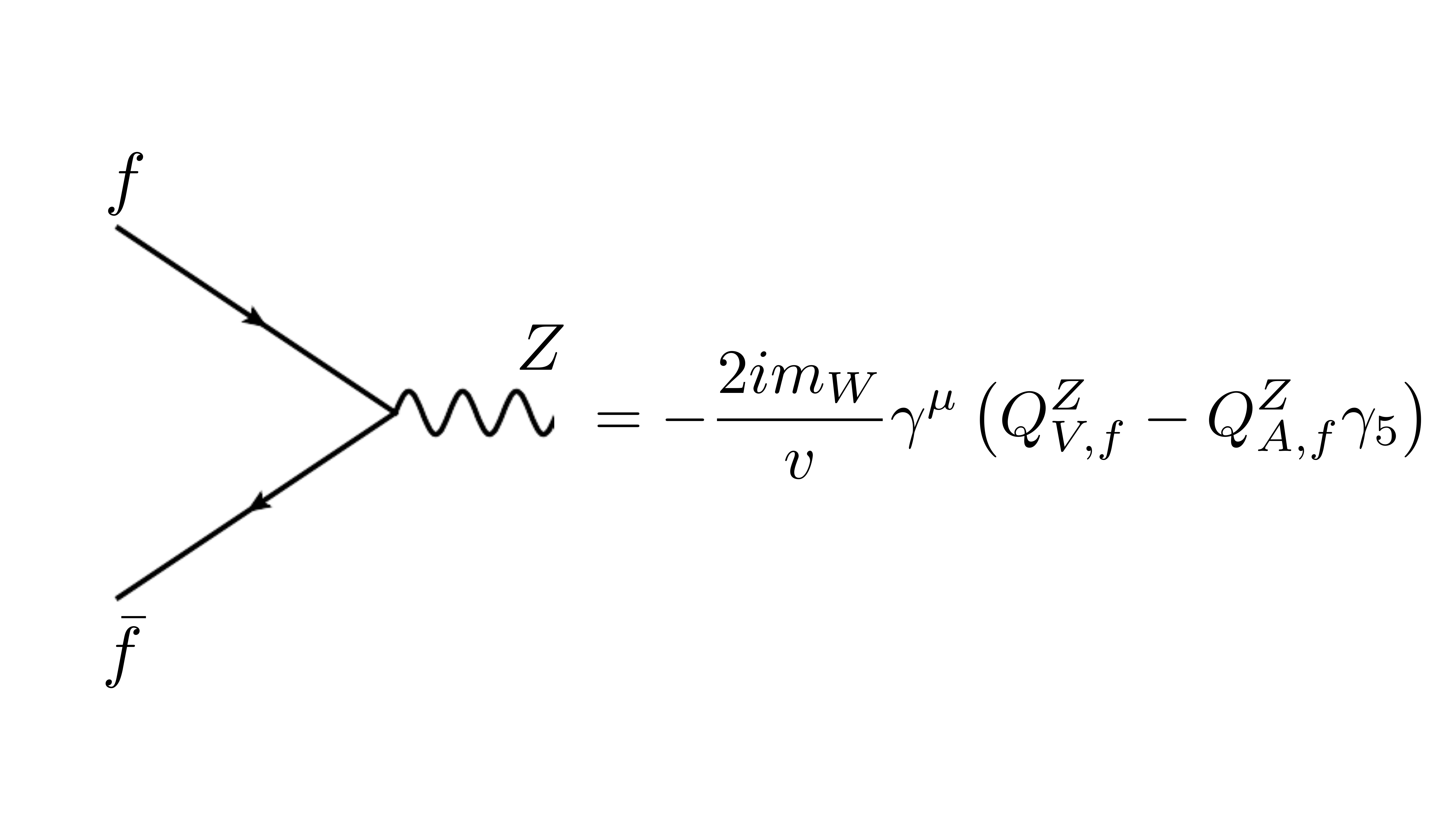}
  \caption{\label{fig:vertex}
Feynman rule coupling a Z boson to fermions.
  }
  \end{center}
  \end{figure}

The coupling of the $Z$-boson to fermions involves a vector and an axial-vector part (see figure~\ref{fig:vertex}).
Accordingly, we may split the partonic coefficient function into two terms accounting for the coupling to the vector and axial-vector parts separately:
\beq
\eta_{\, ij\to B / B^\prime +X} =\hat \sigma\left[\eta^V_{\, ij\to B / B^\prime +X} +\eta^A_{\, ij\to B / B^\prime +X} \right], \hspace{1cm}\hat \sigma=\frac{4 \pi m_W^2}{n_c Q^2 v^2}.
\eeq
The inclusive cross section does not contain any non-zero partonic coefficient functions that depend on both a vector and an axial-vector coupling.
We are interested in the perturbative coefficient functions computed up to a fixed order in the expansion in the strong coupling constant:
\beq
\eta^Y_{\, ij\to B / B^\prime +X} =\sum_{i=0}^\infty a_s(\mu)^i \eta^{Y,\,(i)}_{\, ij\to B / B^\prime +X}, \hspace{1cm} Y\in\{V,A\}\,,\hspace{1cm} a_s(\mu) =\frac{\alpha_S(\mu)}{\pi}\,,
\eeq
where $\alpha_S(\mu)$ denotes the strong coupling constant in the $\overline{\textrm{MS}}$ scheme. At leading order in perturbation theory only the partonic coefficient functions with a quark -- anti-quark pair in the initial state are non-vanishing:
\beq
\eta^{Y,\, (0)}_{q \bar q  \to B / B^\prime}= Q_{Y,q}^{B}Q_{Y,q}^{B^\prime} \delta\left(1-z\right), \hspace{1cm} Y\in\{V,A\}.
\eeq

\begin{figure}[!h]
  \centering
  \begin{subfigure}[b]{0.3\textwidth}
  \includegraphics[width=0.9\textwidth]{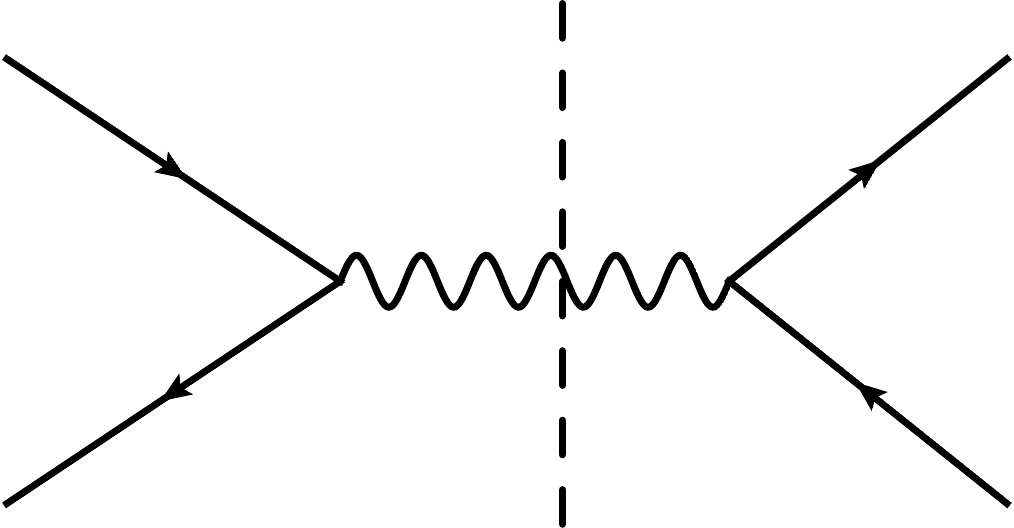}
   \caption{\label{fig:borncharge}$Q_{Y,i}^{B}Q_{Y,j}^{B^\prime} \delta_{ij}$}
  \end{subfigure}
  \begin{subfigure}[b]{0.3\textwidth}
  \includegraphics[width=0.9\textwidth]{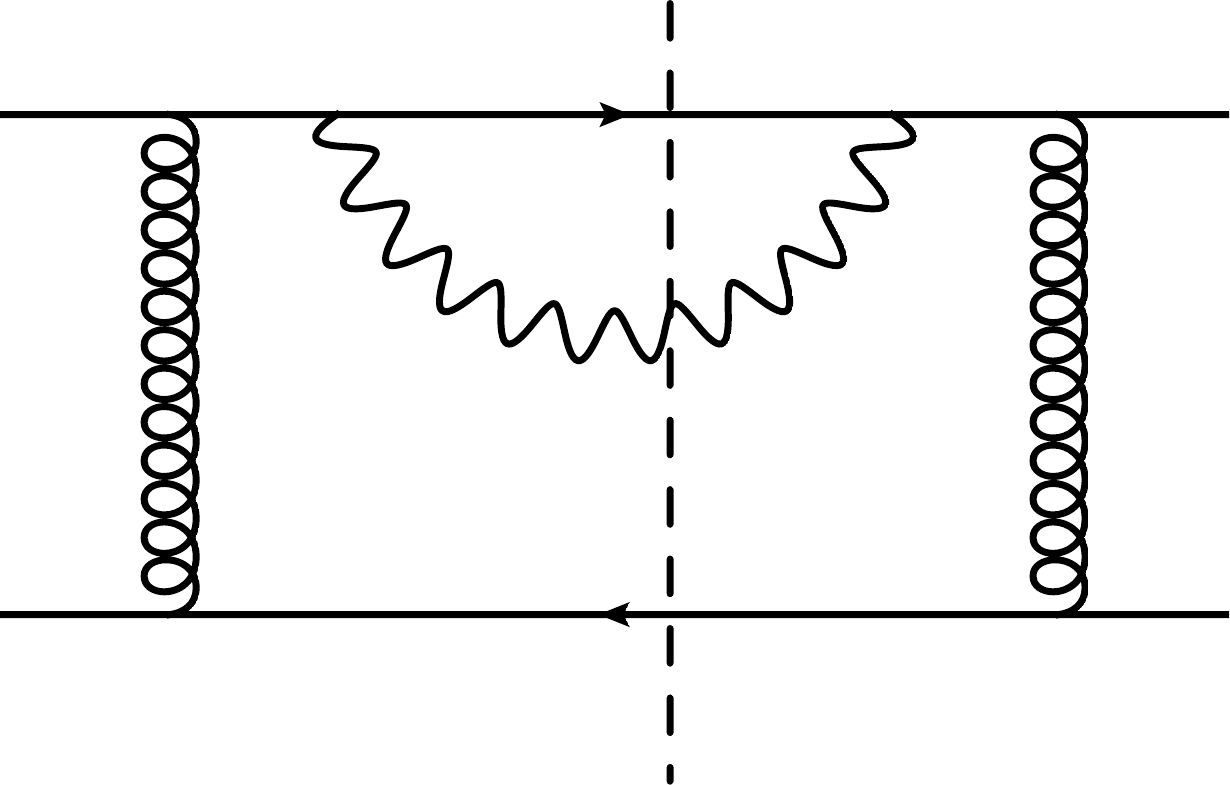}
   \caption{\label{fig:twolineQ2} $Q_{Y,i}^{B}Q_{Y,i}^{B^\prime} $}
  \end{subfigure}
  \begin{subfigure}[b]{0.3\textwidth}
  \includegraphics[width=0.9\textwidth]{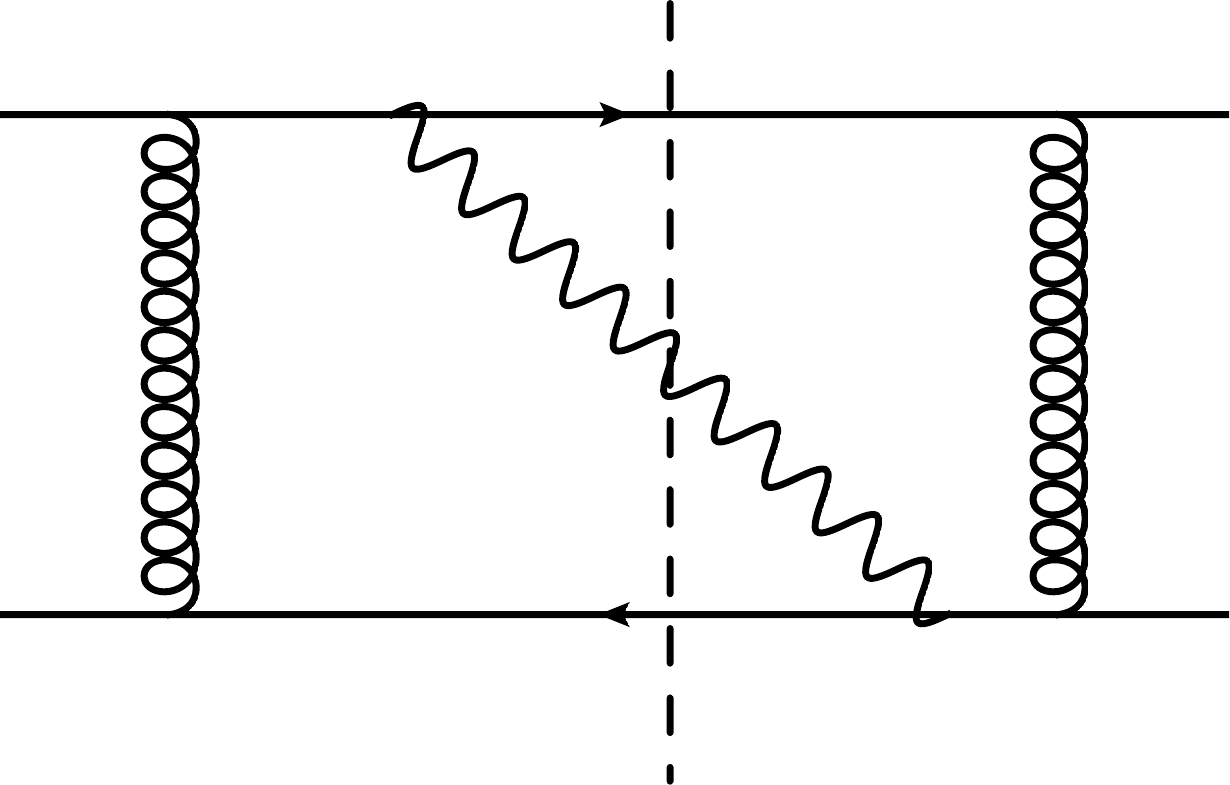}
  \caption{\label{fig:twolineQQ} $Q_{Y,i}^{B}Q_{Y,j}^{B^\prime}$}
  \end{subfigure}  
  \begin{subfigure}[b]{0.3\textwidth}
  \includegraphics[width=0.9\textwidth]{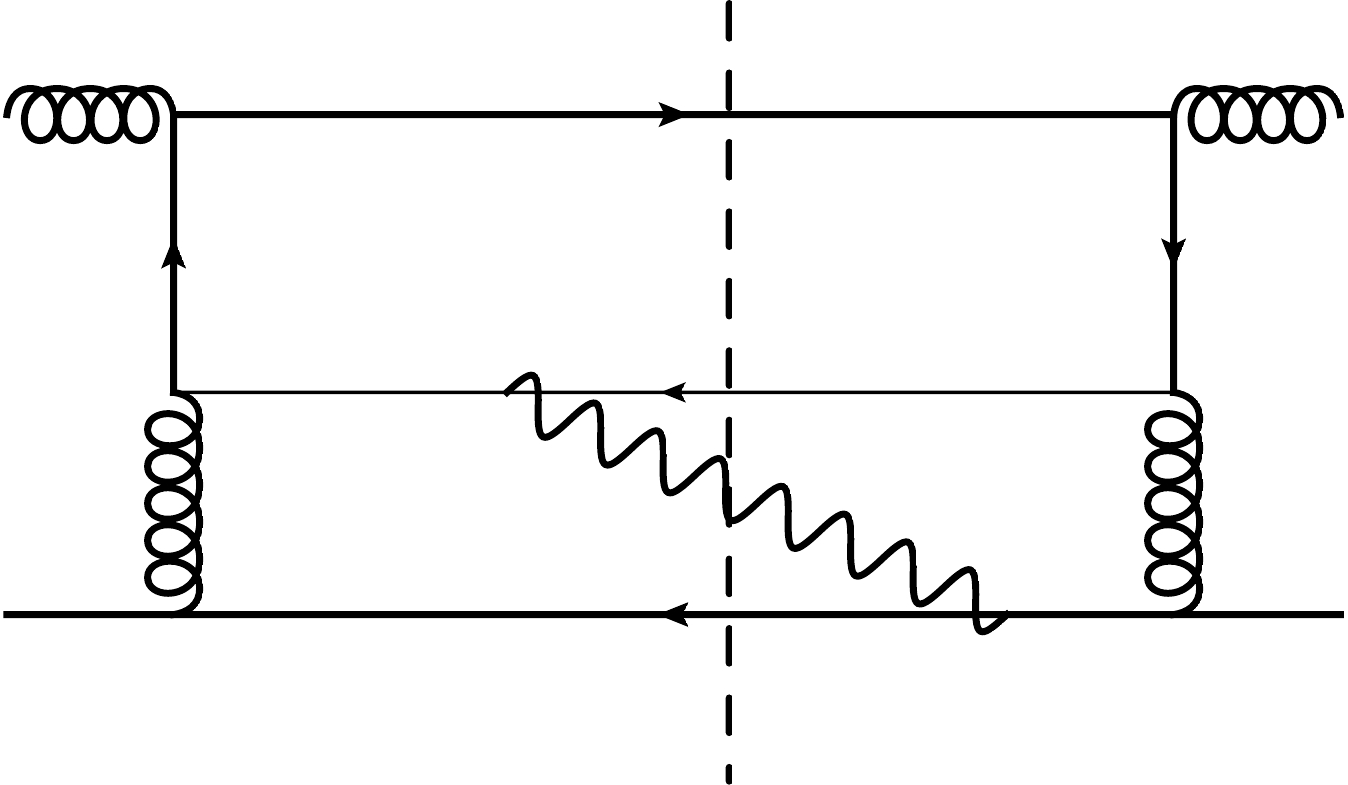}
  \caption{\label{fig:csum}$Q_{Y,j}^{B}\sum_f Q_{Y,f}^{B^\prime}$}
  \end{subfigure}
  \begin{subfigure}[b]{0.3\textwidth}
  \includegraphics[width=0.9\textwidth]{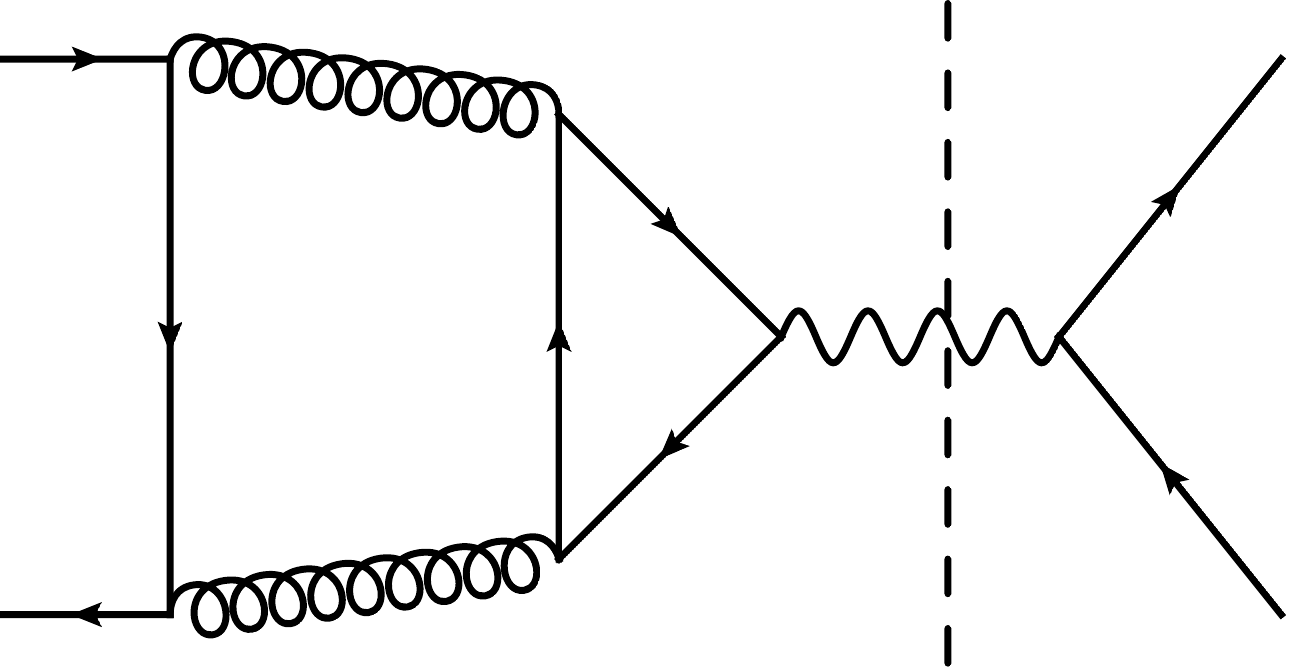}
  \caption{\label{fig:csumdel}$\delta_{ij} Q_{Y,i}^{B}   \sum_f Q_{Y,f}^{B^\prime}  $}
  \end{subfigure}
  \begin{subfigure}[b]{0.3\textwidth}
  \includegraphics[width=0.9\textwidth]{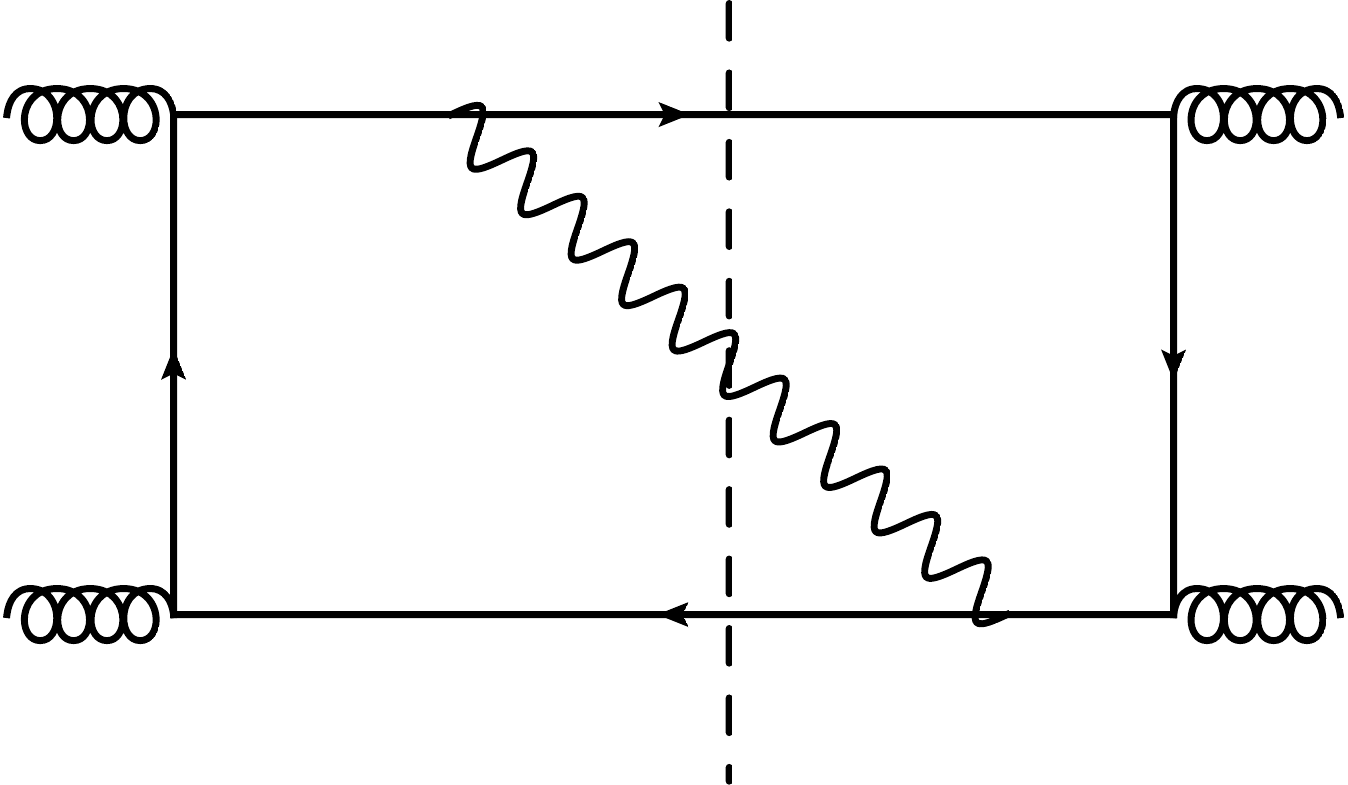}
  \caption{\label{fig:csum2}$\sum_f Q_{Y,f}^{B}Q_{Y,f}^{B^\prime} $}
  \end{subfigure}
  \caption{\label{fig:charges}
Representative diagrams for different charge structures appearing in the NCDY cross section.
The diagrams show interference diagrams where the vertical dashed line represents the final state phase space cut. 
Solid, wavy and curly lines represent propagating quarks, gauge bosons and gluons respectively.
  }
  \end{figure}
  
The partonic coefficient functions for the production of an electroweak gauge boson can be split into 9 different contributions, corresponding to particular combinations of the charges associated with different ways of coupling the electroweak gauge bosons to the quarks. 
For the purpose of simplicity, we write $\eta_{ij} = \eta^Y_{\, ij\to B / B^\prime +X} $. We then have the following decomposition:
\bea
\eta_{ij} &=&
  \eta_{ij} ^{1} \left(\sum_f Q_{Y,f}^{B}\right)\left(\sum_f Q_{Y,f}^{B^\prime}\right)
+ \eta_{ij} ^{2} \sum_f Q_{Y,f}^{B}Q_{Y,f}^{B^\prime} 
+ \eta_{ij}^3 Q_{Y,i}^{B} \sum_f Q_{Y,f}^{B^\prime}
+ \eta_{ij}^4 Q_{Y,j}^{B}\sum_f Q_{Y,f}^{B^\prime}
\nonumber\\
&+&
 \eta_{ij}^5 Q_{Y,i}^{B}Q_{Y,i}^{B^\prime}
+\eta_{ij}^6Q_{Y,j}^{B}Q_{Y,j}^{B^\prime}
+\eta_{ij}^7Q_{Y,i}^{B}Q_{Y,j}^{B^\prime} \delta_{ij}
+\eta_{ij}^8Q_{Y,i}^{B}Q_{Y,j}^{B^\prime}
+\eta_{ij}^9\delta_{ij} Q_{Y,i}^{B}   \sum_f Q_{Y,f}^{B^\prime}  \,.
\label{eq:chargedecomp}
\eea
Figure~\ref{fig:charges} shows representative diagrams that illustrate how the different charge structures arise. 
Figure~\ref{fig:borncharge} shows the tree-level diagram where the initial-state $q\bar{q}$-pair annihilates and produces the gauge boson. 
The flavors of the initial-state quarks are connected by the gauge boson. This is made manifest by the Kronecker delta symbol $\delta_{ij}$ multiplying the corresponding partonic coefficient function $\eta_{ij}^7$.
In contrast, there are classes of diagrams with a quark and anti-quark of the same flavor, but the quark lines are not identical. In such diagrams the gauge boson may or may not connect the different quark lines, as is illustrated in fig.~\ref{fig:twolineQ2} and fig.~\ref{fig:twolineQQ}. Their contributions are included in $\eta_{ij}^5$, $\eta_{ij}^6$ and $\eta_{ij}^8$ respectively. 
Figures~\ref{fig:csum} and~\ref{fig:csumdel} show contributions, denoted by by $\eta_{ij}^3$, $\eta_{ij}^4$ and $\eta_{ij}^9$, where the gauge boson is connected only to one quark line of an initial-state quark.
Furthermore, it is possible that the gauge boson couples to quark lines that are disconnected from the initial state, as shown for example in fig.~\ref{fig:csum2}. These contributions are taken into account by $\eta_{ij}^1$ and $\eta_{ij}^2$.

The partonic coefficient functions $\eta^V_{\, ij\to B / B^\prime +X} $ and $\eta^A_{\, ij\to B / B^\prime +X} $ begin to differ starting from NNLO in QCD perturbation theory. 
At this order, it is for the first time possible to connect two different quark lines with the produced electroweak gauge boson. 
As long as the gauge boson is twice connected to the same quark line and all quarks are treated as massless, helicity conservation dictates that vector and axial-vector partonic coefficient functions will be identical.
In particular, $\eta_{ij}^2$, $\eta_{ij}^5$, $\eta_{ij}^6$, $\eta_{ij}^7$ are identical for vector and axial-vector production at all orders in perturbation theory.

The partonic coefficient functions for the production of a vector boson $\eta^V_{\, ij\to B / B^\prime +X}$ were computed at N$^3$LO in QCD perturbation theory in refs.~\cite{Altarelli:1978id,Altarelli:1979ub,Matsuura:1987wt,Matsuura:1988nd,Matsuura:1988sm,Matsuura:1990ba,Hamberg:1990np,vanNeerven:1991gh,Harlander:2002wh,Duhr:2020seh}.
The computation of the partonic coefficient functions for an axial-vector boson $\eta^A_{\, ij\to B / B^\prime +X}$ through N$^3$LO in perturbative QCD is one of the main results of this article.
We present analytic formul\ae\ for these partonic coefficient functions in terms of ancillary files attached to the arXiv submission of this article. In the next section we discuss in detail the computation and the structure of the partonic coefficient functions for axial-vector production in QCD.


\section{Axial-vector production at N$^3$LO in QCD} 
\label{sec:axial}

Let us consider N$^3$LO corrections to the production cross section for a color-singlet axial-vector state $Z_A$. In broad terms, our computational strategy is similar to the computation of the N$^3$LO corrections to the inclusive Higgs boson production cross sections in gluon fusion~\cite{Anastasiou:2015ema,Mistlberger:2018etf} and bottom-quark fusion~\cite{Duhr:2019kwi}, and the charged-current and photon-only DY processes~\cite{Duhr:2020sdp,Duhr:2020seh}.
More precisely, we generate Feynman diagrams using QGRAF~\cite{Nogueira1993} and perform the spinor and color algebra based on custom C++ algorithms using GiNaC~\cite{Bauer2000} and FORM~\cite{Ruijl:2017dtg}.
We generate all required integrands for interferences of matrix elements for this cross section up to N$^3$LO. 
We then reduce all real and virtual interference diagrams to a set of so-called master integrals using IBP identities~\cite{Tkachov1981,Chetyrkin1981,Laporta:2001dd} and the framework of reverse unitarity~\cite{Anastasiou2002,Anastasiou2003,Anastasiou:2002qz,Anastasiou:2003yy,Anastasiou2004a}. 
The master integrals were computed using the method of differential equations~\cite{Kotikov:1990kg,Kotikov:1991hm,Kotikov:1991pm,Henn:2013pwa,Gehrmann:1999as} in refs.~\cite{Anastasiou:2013mca,Anastasiou:2013srw,Anastasiou:2015yha,Dulat:2014mda,Mistlberger:2018etf}. We work in dimensional regularisation, and compute all matrix elements in $D=4-2\eps$ dimensions.
After inserting the master integrals into our partonic coefficient functions, we renormalise UV divergences in the $\overline{\text{MS}}$ scheme. 
Furthermore, we remove collinear initial-state singularities via mass factorization counterterms comprised of DGLAP~\cite{Gribov:1972ri,Altarelli:1977zs,Dokshitzer:1977sg} splitting functions~\cite{Moch:2004pa,Vogt:2004mw,Ablinger:2014nga,Ablinger:2017tan}.
Finally, we arrive at fully analytic formul\ae\ for our finite partonic coefficient functions for the production of an axial-vector boson through N$^3$LO in perturbative QCD.
The functions are expressed in terms of the class of iterated integrals introduced in ref.~\cite{Mistlberger:2018etf}. While the general strategy is the same as for the N$^3$LO cross section considered in ref.~\cite{Anastasiou:2015ema,Mistlberger:2018etf,Duhr:2019kwi,Duhr:2020sdp}, there are several technical steps that distinguish axial-vector production from those other processes, due to the ambiguities in how to treat $\gamma^5$ in dimensional regularisation. In the remainder of this section we discuss our treatment of $\gamma^5$ in detail.

Consider the operator describing the coupling of an axial current in QCD to the axial-vector $Z_A$:
\beq
\label{eq:ACDef1}
\ord_A= g_A \, Z_{A,\mu}\,\sum_{f=1}^{N_f}A_f\,J_{A,f}^{\mu}\,,\qquad J_{A,f}^{\mu} = i \bar{q}_f\gamma^\mu\gamma^5q_f\,,
\eeq
where $N_f$ is the number of quark species. In the SM we have $g_A = \frac{2m_W}{v\,\cos\theta_W}$, $A_f = \frac{1}{2}T^3_f$ and $N_f=6$. Since the coupling of $Z_A$ to the axial current involves the $\gamma^5$ matrix, which is only well-defined in four dimensions, care is needed how this operator is analytically continued to arbitrary dimensions when working in dimensional regularisation. 
Moreover, it is well-known that the singlet axial current $J_{A,S}^{\mu} = \sum_{f=1}^{N_f}J_{A,f}^{\mu}$ is anomalous in QCD and satisfies the Adler-Bell-Jackiw (ABJ) anomaly equation~\cite{Adler:1969gk,Bell:1969ts,Adler:1969er} (see also ref.~\cite{Bos:1992nd}). 
In four dimensions $\gamma^{\mu}$ and $\gamma^5$ anticommute, as one can easily verify by writing $\gamma_5$ as in refs.~\cite{tHooft:1972tcz,Breitenlohner:1977hr}:
\beq
\gamma^5 = -\frac{i}{4!}\,\eps^{\mu\nu\rho\sigma}\,\gamma_{\mu}\gamma_{\nu}\gamma_{\rho}\gamma_{\sigma}\,.
\eeq
However, if the spacetime dimension is extended to $D$-dimension this no longer holds true.
Here we work in the Larin-scheme~\cite{Larin:1991tj,Larin:1993tq,Jegerlehner:2000dz} and define the axial-vector current explicitly as
\beq
\label{eq:ACDef2}
J_{A,f}^{\mu} = \frac{1}{3!} \epsilon^{\mu\nu\rho\sigma} \bar{q}_f\gamma_\nu \gamma_\rho \gamma_\sigma q_f.
\eeq
For $D=4$, this definition is identical to the one of eq.~\eqref{eq:ACDef1} as on can easily see by computing the anti-commutation relation 
\beq
\gamma^\mu \gamma_5=\frac{1}{2} \{\gamma^\mu,\gamma_5\}=-\frac{i}{3!} \epsilon^{\mu\nu\rho\sigma}\gamma_\nu\gamma_\rho\gamma_\sigma.
\eeq
In our computation the $\gamma^{\mu}$ matrices are interpreted as $D$-dimensional objects, and the Dirac algebra is performed in $D$-dimensions.
Any Feynman diagram in the Drell-Yan process mediated by an axial-vector current will involve two insertions of the axial-vector current of eq.~\eqref{eq:ACDef2}.
We use the identity 
\beq
\epsilon^{\mu_1\mu_2\mu_3\mu_4}\epsilon^{\nu_1\nu_2\nu_3\nu_4} =\det\left(\begin{array}{cccc}
g^{\mu_1\nu_1}&g^{\mu_1\nu_2}&g^{\mu_1\nu_3}&g^{\mu_1\nu_4} \\
g^{\mu_2\nu_1}&g^{\mu_2\nu_2}&g^{\mu_2\nu_3}&g^{\mu_2\nu_4} \\
g^{\mu_3\nu_1}&g^{\mu_3\nu_2}&g^{\mu_3\nu_3}&g^{\mu_3\nu_4} \\
g^{\mu_4\nu_1}&g^{\mu_4\nu_2}&g^{\mu_4\nu_3}&g^{\mu_4\nu_4} \\
\end{array}\right),
\eeq
which is valid in strictly $D=4$ dimensions, to contract Lorentz indices of the Levi-Civita tensors, and we treat the metric tensors in the above equation as $D$-dimensional.
The above extension to $D$ space-time dimensions modifies the computed cross sections at sub-leading order in the dimensional regulator. 
In the presence of divergences, these modifications are propagated into finite and singular terms of the bare partonic cross sections and their renormalisation below.

It is well-known that the axial current $J_{A,f}^{\mu}$ is anomalous in QCD and develops a UV-divergence starting from one-loop order. 
This UV-divergence cancels in the complete SM, due to the relation
\beq\label{eq:anomaly_cancellation}
\sum_{f=1}^{N_f = 6}A_f= 0\,.
\eeq
In variants of the SM with an odd number $N_f$ of fermion species, however, eq.~\eqref{eq:anomaly_cancellation} is violated, and the UV-divergence does not cancel. 
The UV-divergence can be removed by renormalising the axial current by the equation (valid through at least three loops)
\beq\label{eq:axial-renormalisation}
\big[J_{A,f}^{\mu}\big]_R = Z_{ns}\,J_{A,f}^{\mu} + Z_s\,\sum_{f'=1}^{N_f}J_{A,f'}^{\mu}  = Z_{ns}\,J_{A,f}^{\mu} + Z_s\,J_{A,S}^{\mu}\,,
\eeq
where $\big[J_{A,f}^{\mu} \big]_R$ is the renormalised axial current. We see that the renormalisation mixes the axial currents for different flavors. The non-singlet and singlet counterterms $Z_{ns}$ and $Z_s$ are not pure $\overline{\textrm{MS}}$ counterterms, but they also include finite terms
whose purpose is to ensure that the renormalised singlet axial current computed in the Larin-scheme satisfies the all-order ABJ anomaly equation~\cite{Adler:1969gk,Bell:1969ts,Adler:1969er}:
\beq
\partial_{\mu}\big[J_{A,S}^{\mu}\big]_{R} = \frac{\alpha_S(\mu^2)}{8\pi}\,N_f\,\big[F\widetilde{F}\big]_R\,,
\eeq
with $F\widetilde{F} = \eps_{\mu\nu\rho\sigma}\,\textrm{Tr}\big(F^{\mu\nu}F^{\rho\sigma}\big)$, and the renormalised singlet axial current is 
\beq
\big[J_{A,S}^{\mu}\big]_{R}  = Z_S\,J_{A,S}^{\mu}\,, \textrm{~~with~~} Z_S = Z_{ns} + N_f\,Z_s\,.
\eeq
Both the non-singlet and singlet counterterms are known to three loops in QCD~\cite{Larin:1993tq,Ahmed:2015qpa,Ahmed:2021spj,Chen:2021rft}:
\begin{align}
\nonumber Z_{ns}&\,=1-a_s(\mu)\,C_F + a_s(\mu)^2\,C_F\, \left[\frac{11 C_A-2 N_f}{24 \epsilon }+\frac{1}{144} \left(-107 \
C_A+198 C_F+2 N_f\right)\right]\\
\nonumber&\,
+a_s(\mu)^3\,C_F\, \Big\{-\frac{\left(11 C_A-2 N_f\right)^2}{432 \epsilon ^2}+\frac{1}{2592 \epsilon }\big(-416 C_A N_f-2574 C_A C_F+1789 C_A^2\\
\nonumber&\,+360 C_F N_f+4 N_f^2\big)+\frac{1}{5184}\big[(4536 \zeta _3 -6441)\, C_A^2+(17502 -12960 \zeta _3)\, C_A C_F\\
\nonumber&\,+(864 \zeta _3 +356)\, C_A N_f+(7776 \zeta _3 -9990)\, C_F^2-(864 \zeta _3 +186)\, C_F N_f+52 N_f^2\big]\Big\}\\
&\,+\ord(a_s(\mu)^4)\,,\\
\nonumber Z_s &\,= a_s(\mu)^2\,C_F \left(\frac{3}{16 \epsilon }+\frac{3}{32}\right) 
+a_s(\mu)^3\,C_F\Bigg\{\frac{2 N_f-11 C_A}{96 \epsilon ^2}+\frac{109 C_A-162 \
C_F+2 N_f}{576 \epsilon }\\
\nonumber&\,+ \frac{1}{3456}\big[(1404 \zeta _3 -326)\, C_A+(-1296 \zeta _3 +621)\, C_F+176 N_f\big]\Bigg\}+\ord(a_s(\mu)^4)\,.
\end{align}

We have computed the partonic coefficient functions in the Larin-scheme in a variant of the SM with only $N_f=5$ massless active quark flavors, and the top quark is considered infinitely-heavy and thus absent from the computation. If only the strong coupling constant is UV-renormalised, then the partonic coefficient functions still exhibit poles in the dimensional regulator $\eps$ even after mass factorisation. However, all these poles cancel once the axial current is renormalised as in eq.~\eqref{eq:axial-renormalisation}, and we obtain finite results for all partonic coefficient functions.  Besides the explicit analytic cancellation of all ultraviolet and infrared poles, we performed several additional checks to validate our results. 
In particular, we find agreement with our previous computation of ref.~\cite{Duhr:2020seh} of partonic coefficient functions for vector boson production for the part of the partonic coefficient functions that are identical between vector and axial-vector production.
Moreover, purely virtual corrections were computed in ref.~\cite{Ahmed:2021spj,Gehrmann:2021ahy,Gehrmann:2010ue,Gehrmann:2010tu} and we find agreement. Finally, we also checked that our results have the expected dependence on the perturbative scale $\mu=\mu_F=\mu_R$. This check, however, involves a subtle point, which we discuss in detail in the remainder of this section.

Our goal is to compute the coefficient functions $\eta_{ij}^A$, which correspond to the axial-vector contribution from $Z$-boson exchange in the limit where we drop all power-suppressed terms in the top-quark mass. If we denote the partonic coefficient functions obtained from our procedure by $\eta_{ij}^{A,5}$, we find that generically $\eta_{ij}^{A,5}\neq \eta_{ij}^{A}$. To see this, consider the dependence of the coefficient functions on the perturbative scale $\mu := \mu_F=\mu_R$. In the SM, this dependence is governed by the the well-known Dokshitzer-Gribov-Lipatov-Altarelli-Parisi (DGLAP) equation~\cite{Gribov:1972ri,Altarelli:1977zs,Dokshitzer:1977sg}:
\begin{align}\label{eq:SM_RGE}
 \frac{\df}{\df\log\mu^2}\eta^A_{ij} &\,=-P_{ik}\otimes \eta^A_{kj}-P_{jk}\otimes \eta^A_{ik}- \beta\,a_s\,\frac{\partial}{\partial a_s}\eta^A_{ij} \,,
 \end{align}
 where $\beta= -\sum_{n=0}^\infty \beta_n\,a_s^{n+1}$ denotes the QCD $\beta$-function and  $P_{kl}$ is the DGLAP splitting kernel, and we introduced the convolution:
 \beq
 (f\otimes g)(z) = \int_0^1\df x_1\,\df x_2\,\delta(z-x_1x_2)\,f(x_1)\,g(x_2)\,.
 \eeq
 The coefficient functions that we computed, however, satisfy the evolution equation
 \beq\bsp\label{eq:wrong_RGE}
 \frac{\df}{\df\log\mu^2}\eta^{A,5}_{ij} &\,=-P_{ik}\otimes \eta^{A,5}_{kj}-P_{jk}\otimes \eta^{A,5}_{ik}- \beta\,a_s\,\frac{\partial}{\partial a_s}\eta^{A,5}_{ij}\\
 &\,-N_A\,\gamma_{J}\sum_{1\le f_1,f_2\le N_f}(A_{f_1}+A_{f_2})\,\eta^{A,5}_{ij,f_1f_2}
  \,,
\esp\eeq
 where we defined implicitly
 \beq\label{eq:A_decomp}
 \eta^{A,5}_{ij} = \sum_{1\le f_1,f_2\le N_f}A_{f_1}\,A_{f_2}\,\eta^{A,5}_{ij,f_1f_2}\,,
 \eeq
and we find it useful to introduce the quantity $N_A = \sum_{f=1}^{N_f}A_f$.
Note that $N_A=0$ in the complete SM with $N_f=6$ quark species.
$\gamma_{J}$ is the anomalous dimension of the (renormalised) axial current~\cite{Larin:1993tq,Ahmed:2021spj}:
\beq
(-\eps +\beta)\,a_s(\mu)\,{\bf Z}_5^{-1}\frac{\df}{\df\log\mu^2} {\bf Z}_5 = \gamma_J \,{\bf E}_{N_f} + \ord(\eps)\,.
\eeq 
Here we defined the matrix
\beq
{\bf Z}_5 = Z_{ns}\, {\bf I}_{N_f} + Z_s\,{\bf E}_{N_f}\,,
\eeq 
where ${\bf I}_{N_f}$ is the $N_f\times N_f$ unit matrix and ${\bf E}_{N_f}$ is the $N_f\times N_f$ matrix whose entries are all 1. We have
 \beq\bsp\label{eq:gamma_A}
\gamma_{J} 
&\,=-a_s(\mu)^2 \,N_A\,\frac{3C_F}{8} - a_s(\mu)^3\,N_A\,\frac{C_F}{8}\,\Bigg(\frac{71 }{12}\,C_A-\frac{9}{4 }\,C_F-\frac{N_f}{6}\Bigg)+\ord(a_s(\mu)^4)\,.
\esp\eeq
 Clearly, the term proportional to $\gamma_J$ in eq.~\eqref{eq:wrong_RGE} is absent in the complete SM, and so $\eta^{A,5}_{ij}$ cannot be identified with the large-$m_t$ limit of the SM. Note that in the complete SM we have $N_A=0$, and so $\gamma_{J} =0$.
 
 The interpretation and resolution of this puzzle is as follows: If we start from the complete SM, with $N_f=6$ and $m_t<\infty$, the UV-divergences from the axial anomaly cancel. Naively, one would expect that by making the top quark infinitely heavy, it decouples from the theory and we land on an effective field theory with $N_f=5$ massless quarks, defined by simply ignoring operators involving the top quark. This naive expectation, however, is wrong: Since the UV divergences from the axial current do not cancel for $N_f=5$, our EFT is anomalous and the top quark does not naively decouple (see, e.g., ref.~\cite{Collins:1978wz}). We include non-decoupling effects in the form of a Wilson coefficient $C_{A,f}(\mu) = A_{f} + \ord(a_s^2)$:
 \beq
 A_f\,J_{A,f}^{\mu} = i A_{f}\,\bar{q}_f\gamma^\mu\gamma^5q_f \to  J_{A,f,\textrm{eff}}^{\mu} = i C_{A,f}(\mu)\,\bar{q}_f\gamma^\mu\gamma^5q_f \,.
 \eeq
The Wilson coefficient satisfies a renormalisation group equation which precisely cancels the contribution from the anomalous dimension $\gamma_{J}$ in eq.~\eqref{eq:wrong_RGE}:
\beq\label{eq:C5_RGE}
 \frac{\df}{\df\log\mu^2}C_{A,f}(\mu) = -\gamma_J\,\sum_{f'=1}^{N_f}C_{A,f'}(\mu)\,.
 \eeq
 We see that the partonic coefficient functions $\eta^{A}_{ij}$ in the large-$m_t$ limit of the SM are obtained from $\eta^{A,5}_{ij}$ by the replacement $ A_{f} \to  C_{A,f}(\mu)$ (cf. eq.~\eqref{eq:A_decomp}):
 \beq\label{eq:eta_A_def}
 \eta_{ij}^A = \sum_{1\le f_1,f_2\le N_f}C_{A,f_1}(\mu)\,C_{A,f_2}(\mu)\,\eta^{A,5}_{ij,f_1f_2}\,.
 \eeq
  The Wilson coefficient admits the perturbative expansion:
 \beq\label{eq:C5_expansion}
C_{A,f}(\mu) = A_f + N_A\,\sum_{n=2}^\infty \sum_{k=0}^{n-1}a_s(\mu)^nc_{A}^{(n,k)}\,\log^k\frac{m_t^2}{\mu^2}\,.
\eeq
It is easy to check that by combining the evolution equations~\eqref{eq:wrong_RGE} and~\eqref{eq:C5_RGE}, the function $\eta_{ij}^A$ defined in eq.~\eqref{eq:eta_A_def} satisfies the SM evolution equation~\eqref{eq:SM_RGE}.

The Wilson coefficient depends explicitly on the (on-shell) top-quark mass, making the non-decoupling of the top quark manifest. It was recently computed through three loops in ref.~\cite{Ju:2021lah,Chen:2021rft}. We have independently obtained all the coefficients except for the non-logarithmic three-loop  coefficient $c^{(3,0)}_A$, and we find full agreement. Indeed, the coefficients of the logarithms, i.e., the $c_{A}^{(n,k)}$ with $k\neq 0$, are determined from the evolution equation~\eqref{eq:C5_RGE}. The non-logarithmic terms $c_{A}^{(n,0)}$ are not constrained by the evolution equation and need to be computed explicitly. We have recalculated the two-loop coefficient $c^{(2,0)}_A$ by taking the large-$m_t$ limit on the known NNLO cross section in the full SM, including all finite quark-mass effects~\cite{Dicus:1985wx,Gonsalves:1991qn,Rijken:1995gi}. The two-loop coefficients are:
\beq\bsp
c_{A}^{(2,1)} &\,= -\frac{3}{8}\,C_F \qquad \textrm{~~and~~} \qquad c_{A}^{(2,0)} = -\frac{3}{16}\,C_F\,\,.
\esp\eeq
At three-loops, we have~\cite{Ju:2021lah,Chen:2021rft}:
\beq\bsp
 c_{A}^{(3,2)} &\,= \frac{3}{8}\,C_F\,\beta_0 = \frac{C_F}{32}(11\,C_A-2\,N_f)\,,\\
   c_{A}^{(3,1)} &\,= -\frac{C_F}{96}\,(38 C_A-27 C_F+4N_f)\,,\\
c_{A}^{(3,0)}&\, = \frac{C_F}{1152}\big[C_A\,(1649-1512\,\zeta_3)+27 C_F\,(48\zeta_3+17)-374N_f-328\big]\,.
 \esp\eeq


\section{Phenomenological Results}
\label{sec:pheno}

In this section we present for the first time phenomenological results for the invariant-mass distribution of a massless lepton pair at N$^3$LO in QCD,
\beq
\Sigma^{\textrm{N$^k$LO}}(Q^2)= Q^2\,\frac{\df\sigma^{\textrm{N$^k$LO}}}{\df Q^2}\,.
\eeq
We work in the five-flavour scheme with $N_f=5$ active massless quark flavors. The top-quark is considered infinitely heavy, though the non-decoupling effects are included through the Wilson coefficient $C_{A,f}$. We mention that finite quark-mass effects have been calculated through NNLO and are known to be very small~\cite{Dicus:1985wx,Gonsalves:1991qn,Rijken:1995gi}. We therefore expect our computation to give a very good estimate of the QCD corrections in the full SM where the full finite top-mass effects are retained. Our choice for the numerical values of the SM input parameters is summarized in table~\ref{tab:input}. The strong coupling constant is evolved from $\alpha_S(m_Z)$ to the renormalisation scale $\mu_R$ using the four-loop QCD beta function~\cite{VanRitbergen:1997va,Czakon:2004bu,Baikov:2016tgj,Herzog:2017ohr} in the $\overline{\textrm{MS}}$-scheme using $N_f=5$ active massless quark flavors.
Unless stated otherwise, all results are obtained for a proton-proton collider with $\sqrt{S} = 13$ TeV using the zeroth
member of the combined \verb+PDF4LHC15_nnlo_mc+ set~\cite{Butterworth:2015oua}, and bands correspond to varying the perturbative scales $\mu_F$ and $\mu_R$ by a factor of two around the central scale $\mu_F=\mu_R=Q$ while respecting the constraint
\beq
\frac{1}{2}\leq \frac{\mu_R}{\mu_F}\leq 2.
\eeq
Commonly, this is referred to as 7-point variation.

\begin{table}[!h]
\begin{center}
\begin{tabular}{ccccccc}
\hline\hline
$\alpha_{EW}^{-1}$ & $\alpha_S(m_Z)$ & $v$ [GeV] & $m_Z$ [GeV] & $\Gamma_Z$ [GeV] & $m_W$ [GeV] & $m_t$ [GeV] \\
\hline
132.186 & 0.118 & 246.221 & 91.1876  & 2.4952 & 80.379 & 172.9\\
\hline\hline
\end{tabular}
\caption{\label{tab:input} Numerical values of the SM input parameters used for our phenomenological predictions.}
\end{center}
\end{table}

\subsection{Scale dependence and perturbative convergence}

In table~\ref{tab:xsecs} we show values for a selection of representative invariant masses $Q^2$ for the choice of central scale $\mu_F=\mu_R=Q$, together with the corresponding QCD K-factors:
\beq
\textrm{K}^{\text{N$^3$LO}}(Q^2) = \frac{\Sigma^{\textrm{N$^3$LO}}(Q^2)}{\Sigma^{\textrm{NNLO}}(Q^2)}\,.
\eeq
We include an estimate of the residual perturbative uncertainty based on seven point variation of the factorization and renormalisation scale, as well as the uncertainties from PDFs and the value of the strong coupling constant (see below).

\begin{table}[!h]
\begin{center}
\begin{tabular}{c | c | c | c | c }
\hline\hline
$Q${ [GeV]} &  $\Sigma^{\text{N3LO}}$ { [pb]} & $\textrm{K}^{\text{N$^3$LO}}$ & $\delta(\text{PDF-$\alpha_S$})$ & $\delta(\text{PDF-TH})$  \\
\hline
30 & $531.7^{+1.53\%}_{-2.54 \%}$ & 0.952 & ${}^{+3.7 \%}_{-3.8\%}$ & 2.8\% \\
60 & $112.636^{+0.97\%}_{-1.29 \%}$  & 0.97 & ${}^{+2.8 \%}_{-2.5\%}$ & 2.5\% \\
91.1876 & $21756.4^{+0.7\%}_{-0.86 \%} $& 0.977 & ${}^{+2.2 \%}_{-2.1\%}$ &  2.5\% \\
100 & $458.473^{+0.66\%}_{-0.79 \%}$  & 0.979 & ${}^{+2.0 \%}_{-1.8\%}$ & 2.5\%  \\
300 & $1.24661^{+0.26\%}_{-0.29 \%}$  & 0.992 & ${}^{+1.9 \%}_{-1.6\%}$ & 1.7\% \\
\hline\hline
\end{tabular}
\caption{\label{tab:xsecs}Cross section values for a selection of representative invariant mass $Q$ for the choice of central scale $\mu_F=\mu_R=Q$, together with the corresponding QCD K-factors and uncertainties (see main text for a detailed description).}
\end{center}
\end{table}

\begin{figure*}[!h]
\centering
\includegraphics[width=0.95\textwidth]{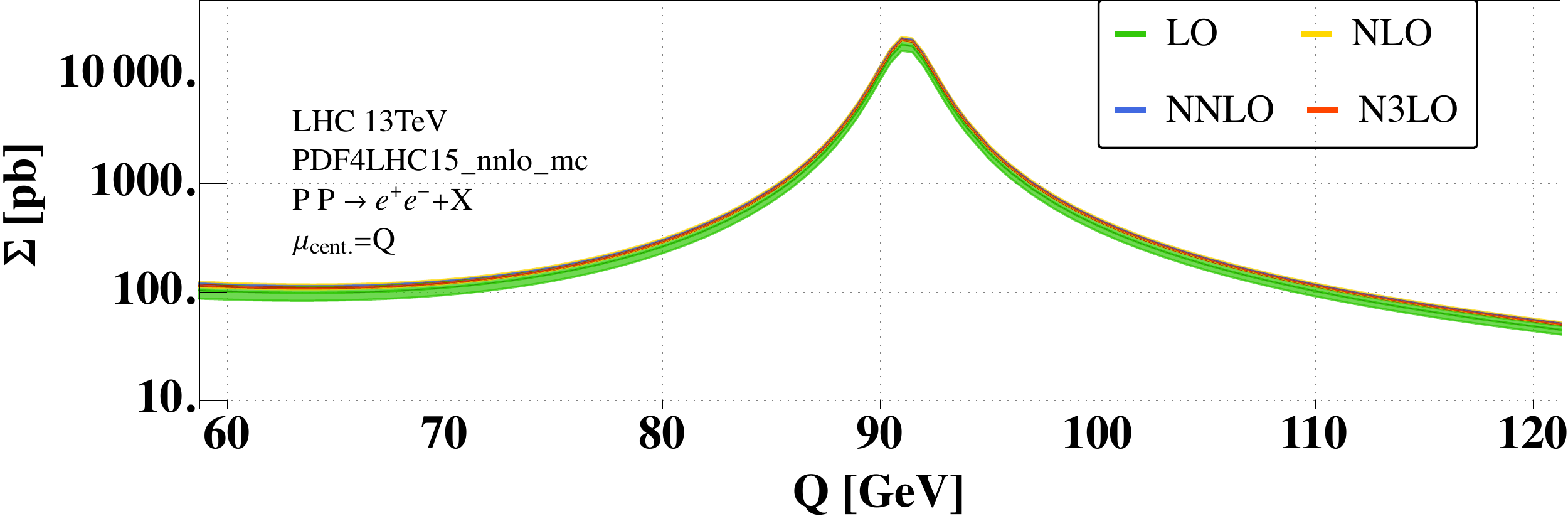}
\caption{\label{fig:totalxs}The invariant-mass distribution $\Sigma(Q^2)$ at the LHC with $\sqrt{S}=13$ TeV at different orders in perturbation theory. 
}
\end{figure*}

In figure~\ref{fig:totalxs} we show the inclusive cross section for the production of a massless lepton pair as a function of $Q^2$.

\begin{figure*}[!h]
\centering
\includegraphics[width=0.95\textwidth]{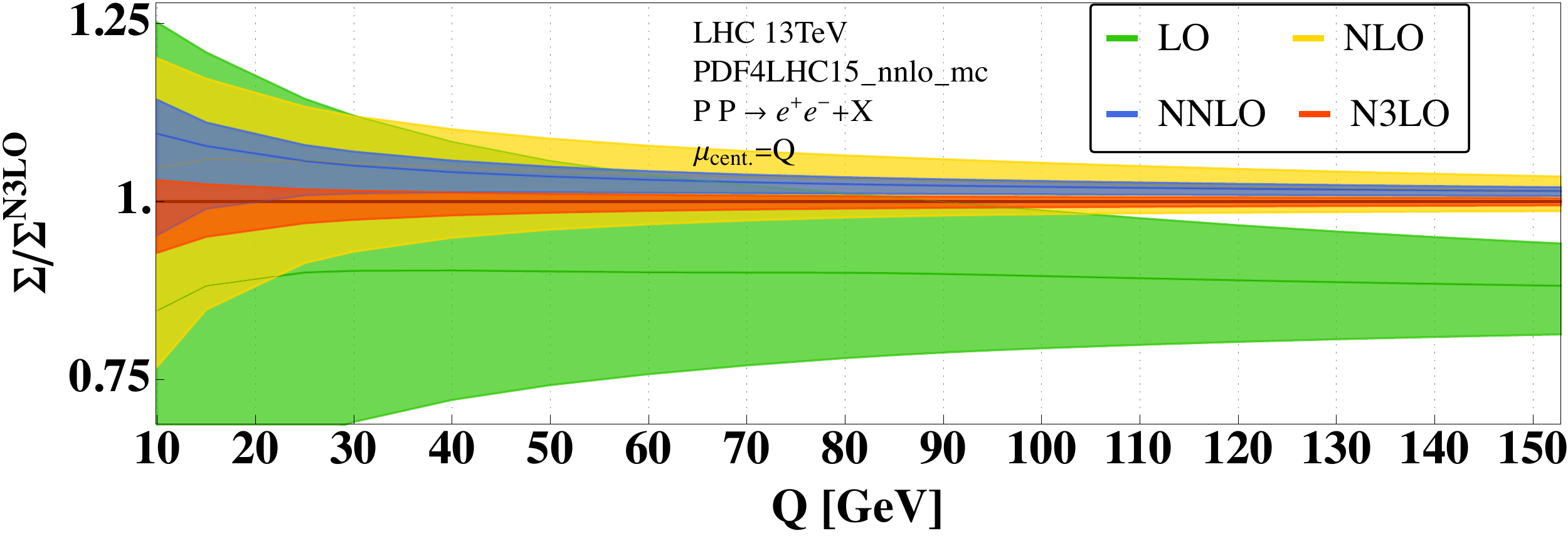}
\caption{\label{fig:KfacmallQ}
The K-factors $\Sigma^{\textrm{N$^k$LO}}/\Sigma^{\textrm{N$^3$LO}}$ as a function of invariant masses 10 GeV$\le Q\le$150 GeV for $k\le 3$. The bands are obtained by varying the perturbative scales by a factor of two around the central $\mu_{\textrm{cent.}} = Q$.
}
\end{figure*}

\begin{figure*}[!h]
\centering
\includegraphics[width=0.95\textwidth]{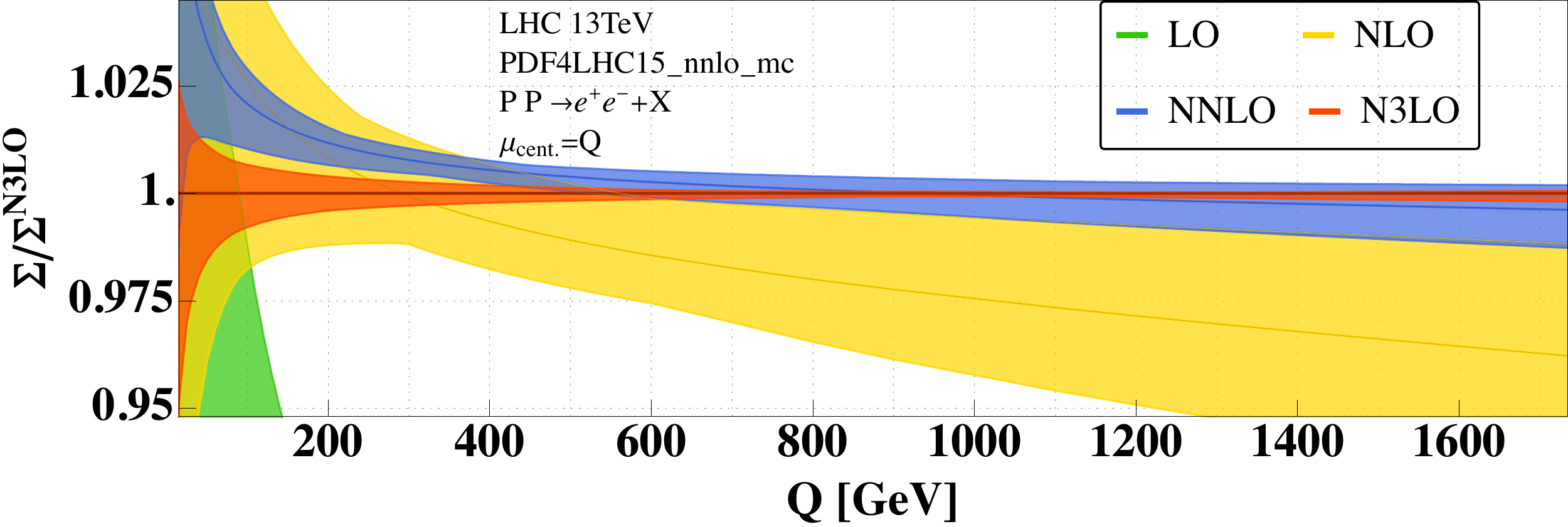}
\caption{\label{fig:KfaclargeQ}
The K-factors $\Sigma^{\textrm{N$^k$LO}}/\Sigma^{\textrm{N$^3$LO}}$ as a function of invariant masses $ Q\le$1.800 GeV for $k\le 3$. The bands are obtained by varying the perturbative scales by a factor of two around the central $\mu_{\textrm{cent.}} = Q$.
}
\end{figure*}

In figures~\ref{fig:KfacmallQ} and~\ref{fig:KfaclargeQ} we show the NCDY cross section normalized to its value computed at N$^3$LO as a function of the invariant mass $Q$. 
We observe qualitatively the same features as for the photon-only~\cite{Duhr:2020seh,Chen:2021vtu} and the charged-current DY processes~\cite{Duhr:2020sdp}. 
Specifically, we observe that in the range of invariant masses between $\sim 40$ GeV and $\sim 400$ GeV, the scale variation bands from NNLO and N$^3$LO do not overlap, indicating that conventional scale variation at NNLO underestimated the true size of the N$^3$LO corrections. 
We note, however, that the size of the bands at NNLO was particularly small for the NCDY process, often at the sub-percent level depending on the invariant masses considered.

\begin{figure*}[!h]
\centering
\includegraphics[width=0.45\textwidth]{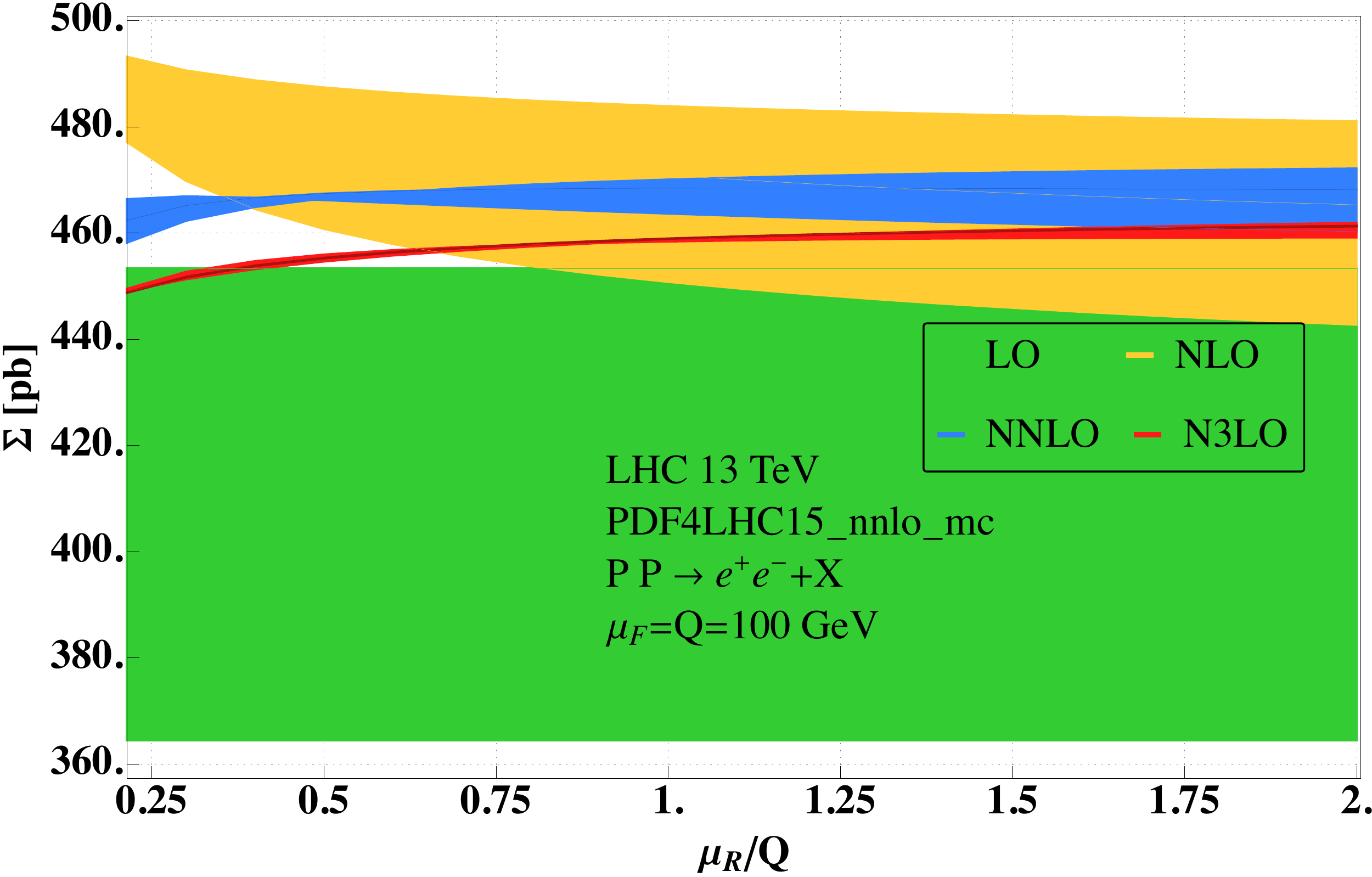}
\includegraphics[width=0.45\textwidth]{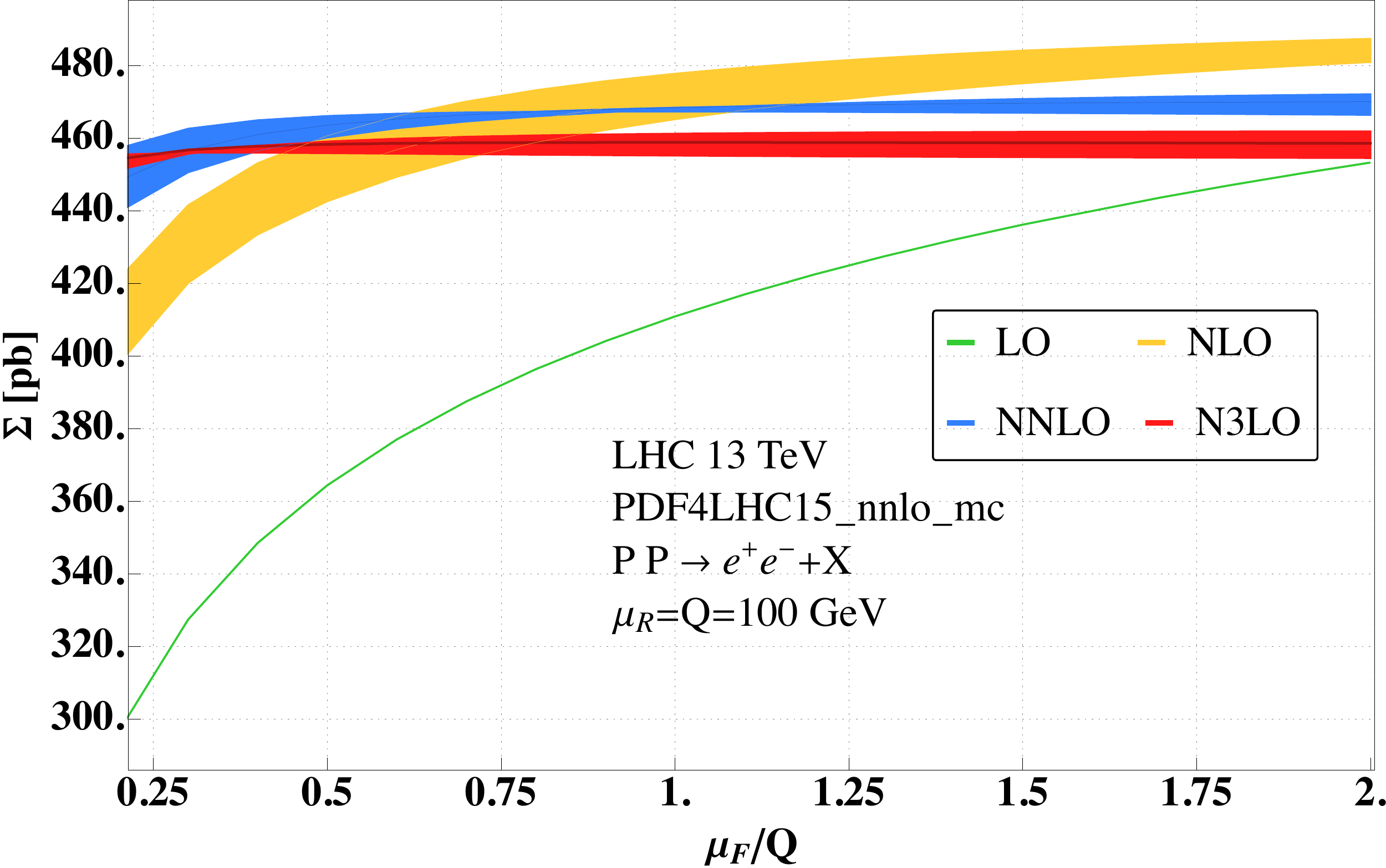}
\caption{\label{fig:scalevariation}
Dependence of the invariant-mass distribution through N$^3$LO on one of the two perturbative scales with the other held fixed. The bands are obtained by varying the other scale by a factor of two around the central scale $Q=100$ GeV.}
\end{figure*}

In figure~\ref{fig:scalevariation} we show the dependence of the cross section for $Q=100$ GeV on one of the two perturbative scales with the other held fixed at some value in the interval $[Q/2,2Q]$. We observe a very good reduction of the scale dependence as we increase the perturbative order, with only a very mild scale dependence at N$^3$LO. Just like for the photon-only and $W$ cases, the bands from NNLO and N$^3$LO do not overlap.~\footnote{The leading order cross section does not depend on the strong coupling constant and consequently does also not change with variation of the renormalisation scale. As a result the right panel of fig.~\ref{fig:scalevariation} does not show any band for the leading order cross section.}

\subsection{Ratios of K-factors}
\label{sec:ratios}
In the previous section we have seen that the NCDY process shows relatively good perturbative stability as we increase the order.
This behavior is very reminiscent of the photon-only and charged-current DY processes considered in refs.~\cite{Duhr:2020seh,Duhr:2020sdp}. 
To investigate if and to what degree QCD corrections differ among different DY-type processes, we study ratios of K-factors between the NCDY process and the photon-only and charged-current DY processes. 
In particular, in this section we refer to the ratio of the N$^n$LO cross section to its LO counter part as the N$^n$LO K-factor, K$^{(n)}$.
In contrast, ratios of cross sections rather than K-factors involving the $W$ and $\gamma^*$ cross sections were considered in ref.~\cite{Duhr:2020sdp}. 

Before we discuss these ratios, we need to make a comment. Whenever one studies ratios of cross sections, there is an ambiguity in how to choose the perturbative scales. For example, if one believes that QCD corrections should be similar between $W$, $\gamma^*$ and $Z$ production (as motivated for example by the universality of certain limits, like the threshold limit), it is natural to vary the scales in the numerator and the denominator in a correlated way. Alternatively, one may choose the scales in an uncorrelated way (e.g., because different partonic channels are weighted differently by the PDFs for these processes, breaking the universality of the QCD corrections), typically leading to larger scale variation bands. In 
ref.~\cite{Duhr:2020sdp} it was shown that for the $W$ and $\gamma^*$ cross sections the correlated prescription leads to a vanishingly small scale dependence at the sub-permille level at N$^3$LO, while the uncorrelated prescription produces excessively large scale variation bands at N$^3$LO (much larger than the absolute shift from NNLO to N$^3$LO). As a consequence, neither the correlated nor the uncorrelated prescriptions are expected to give reliable estimates for missing higher-order terms for these ratios at N$^3$LO. Therefore, in ref.~\cite{Duhr:2020sdp} a new prescription was considered, which uses the relative size of the last considered order compared to
the previous one as an estimator of the perturbative uncertainty:
\beq
\delta(\text{pert.})=\pm\left|1-\frac{\textrm{K}^{(n)}_X/\textrm{K}^{(n)}_Y}{\textrm{K}^{(n-1)}_X/\textrm{K}^{(n-1)}_Y}\right|\times100 \%\,.
\eeq
In the following we use this prescription to obtain uncertainty bands for the ratios.

\begin{figure*}[!h]
\centering
\includegraphics[width=0.95\textwidth]{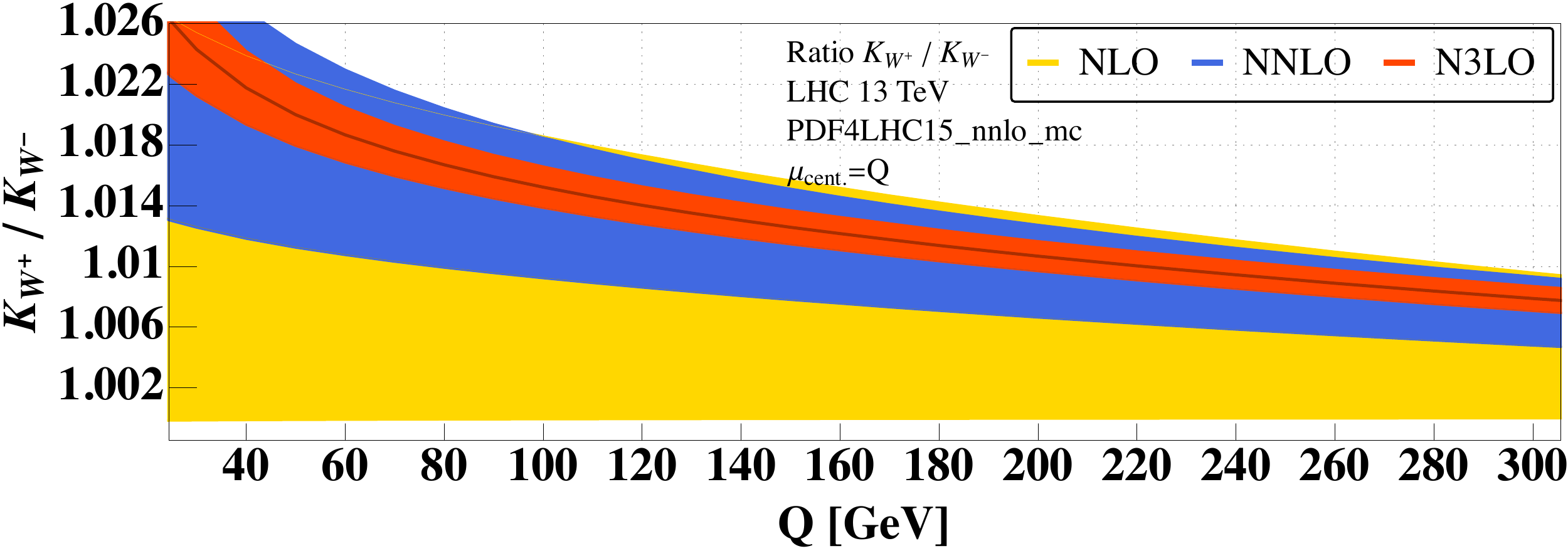}
\caption{\label{fig:KW+/KW-}
Ratio of the K-factors for $W^+$ and $W^-$ production as a function of $Q^2$ at different orders in perturbation theory. 
}
\end{figure*}

\begin{figure*}[!h]
\centering
\includegraphics[width=0.95\textwidth]{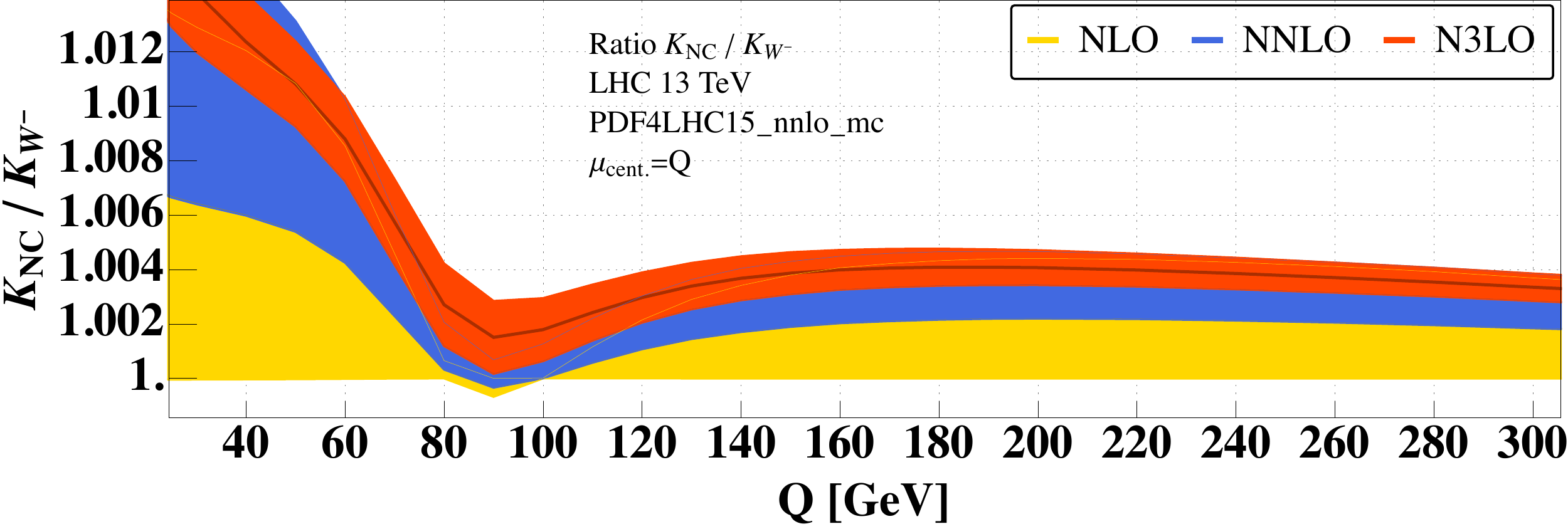}
\caption{\label{fig:KNC/KW-}
Ratio of the K-factors for NCDY and $W^-$ production as a function of $Q^2$ at different orders in perturbation theory. 
}
\end{figure*}

\begin{figure*}[!h]
\centering
\includegraphics[width=0.95\textwidth]{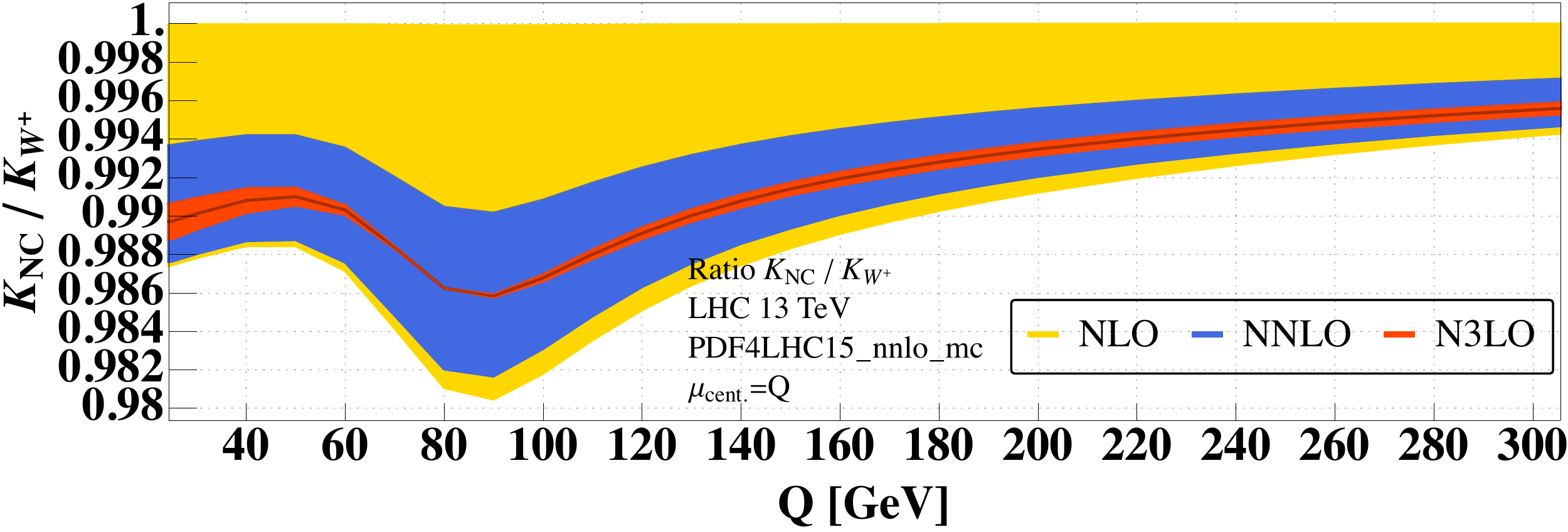}
\caption{\label{fig:KNC/KW+}
Ratio of the K-factors for NCDY and $W^+$ production as a function of $Q^2$ at different orders in perturbation theory. 
}
\end{figure*}

\begin{figure*}[!h]
\centering
\includegraphics[width=0.95\textwidth]{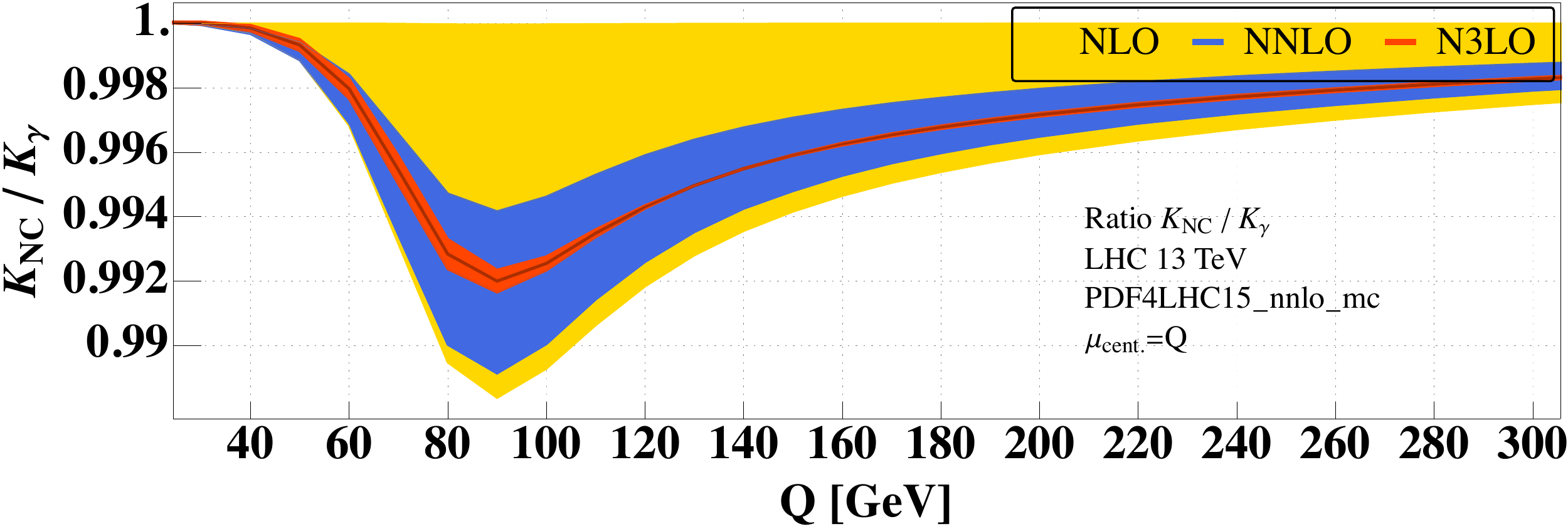}
\caption{\label{fig:KNC/Kphoton}
Ratio of the K-factors for NCDY and $\gamma^*$ production as a function of $Q^2$ at different orders in perturbation theory. 
}
\end{figure*}

In figures~\ref{fig:KW+/KW-} -- \ref{fig:KNC/Kphoton} we show the ratios of the K-factors as a function of the invariant mass $Q$. In all cases we observe a remarkable similarity of the K-factors, which agree among themselves within at worst 2\% for all ratios considered. In particular, we see from figure~\ref{fig:KNC/Kphoton} that the K-factors agree within $1\%$ for the NCDY process computed with or without including the contributions from the $Z$ boson. 
At the same time, we observe that there is a dependence of the shape of the QCD corrections on the invariant mass, and different invariant-mass regions may receive slightly different QCD corrections. 
This shows that, if we want to reach a level of precision of 1\%, care is needed when using K-factors obtained from one process in one region of phase space to reweight other processes or other regions of phase space.
However, our results also demonstrate that shape differences mainly introduced at lower orders in perturbation theory and higher order corrections are remarkably similar.

\subsection{Uncertainties related to PDFs}

In order to assess the dependence of our predictions on the methodology of how the PDFs are extracted, we follow the prescription of ref.~\cite{Butterworth:2015oua} for the computation of PDF uncertainties $\delta(\text{PDF})$ using the Monte Carlo method.
The PDF set $\mathtt{PDF4LHC15\_nnlo\_mc}$ uses $\alpha_S=0.118$ as a central value and two additional PDF sets are available that allow for the correlated variation of the strong coupling constant in the partonic cross section and the PDF sets to $\alpha_S^{\text{up}}=0.1195$ and $\alpha_S^{\text{down}}=0.1165$. 
These sets allow us to deduce an uncertainty $\delta(\alpha_S)$ on our cross section following the prescription of ref.~\cite{Butterworth:2015oua}.
We combine the PDF and strong coupling constant uncertainties in quadrature to give
\beq
\delta(\text{PDF}+\alpha_S)=\sqrt{\delta(\text{PDF})^2+\delta(\alpha_S)^2}\,.
\eeq

Currently there is no available PDF set extracted from data with N$^3$LO accuracy, and so we are bound to use NNLO PDFs in our predictions.  We estimate the potential impact of this mismatch on our results using the prescription
introduced in ref.~\cite{Anastasiou:2016cez}. The PDF theory (PDF-TH) uncertainty is then obtained by studying the variation of the NNLO cross section as NNLO- or NLO-PDFs are used: 
\beq
\label{eq:PDFTH}
\delta(\text{PDF-TH})=\frac{1}{2}\left|\frac{\Sigma^{\text{NNLO, NNLO-PDFs}}(Q^2)-\Sigma^{\text{NNLO, NLO-PDFs}}(Q^2)}{\Sigma^{\text{NNLO, NNLO-PDFs}}(Q^2)}\right|.
\eeq
Here, the factor $\frac{1}{2}$ is introduced as it is expected that this effect becomes smaller at N$^3$LO compared to NNLO.

\begin{figure*}[!h]
\centering
\includegraphics[width=0.95\textwidth]{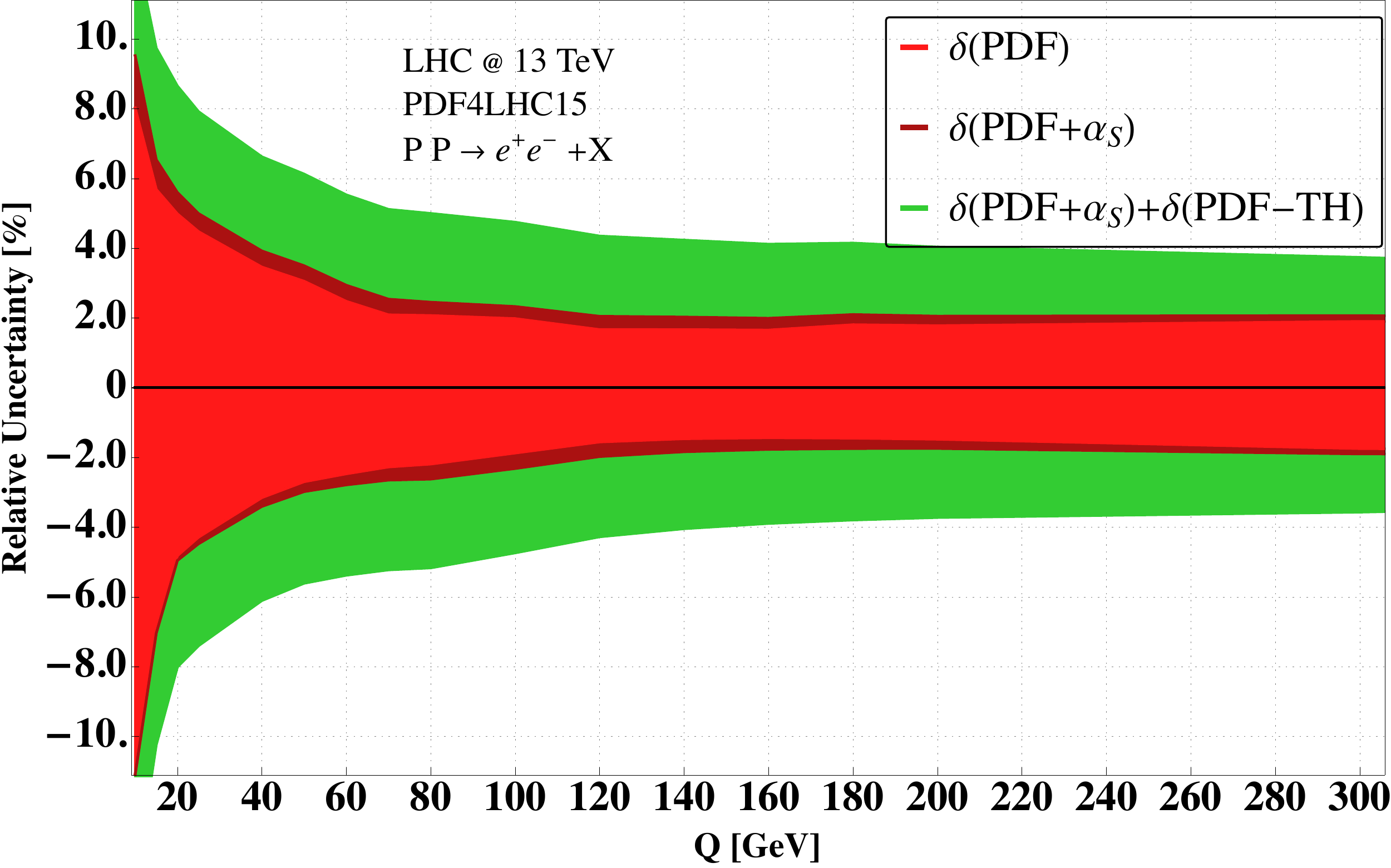}
\caption{\label{fig:PDF_uncert}
Relative uncertainty of the NCDY process at N$^3$LO due to incomplete knowledge of parton distribution functions and the strong coupling constant as a function of $Q$. 
$\delta(\textrm{PDF})$, $\delta(\textrm{PDF}+\alpha_s)$ and the sum of $\delta(\textrm{PDF}+\alpha_s)$ and $\delta(\textrm{PDF-TH})$ are shown in red, brown and green respectively.
}
\end{figure*}
In figure~\ref{fig:PDF_uncert} we show the combined uncertainty from PDFs, the value of the strong coupling constant $\alpha_S$ and the missing N$^3$LO PDFs. The size of these uncertainties is  comparable to the uncertainties obtained in refs.~\cite{Duhr:2020seh,Duhr:2020sdp} for the photon-only and charged-current DY processes.

\begin{figure*}[!h]
\centering
\includegraphics[width=0.95\textwidth]{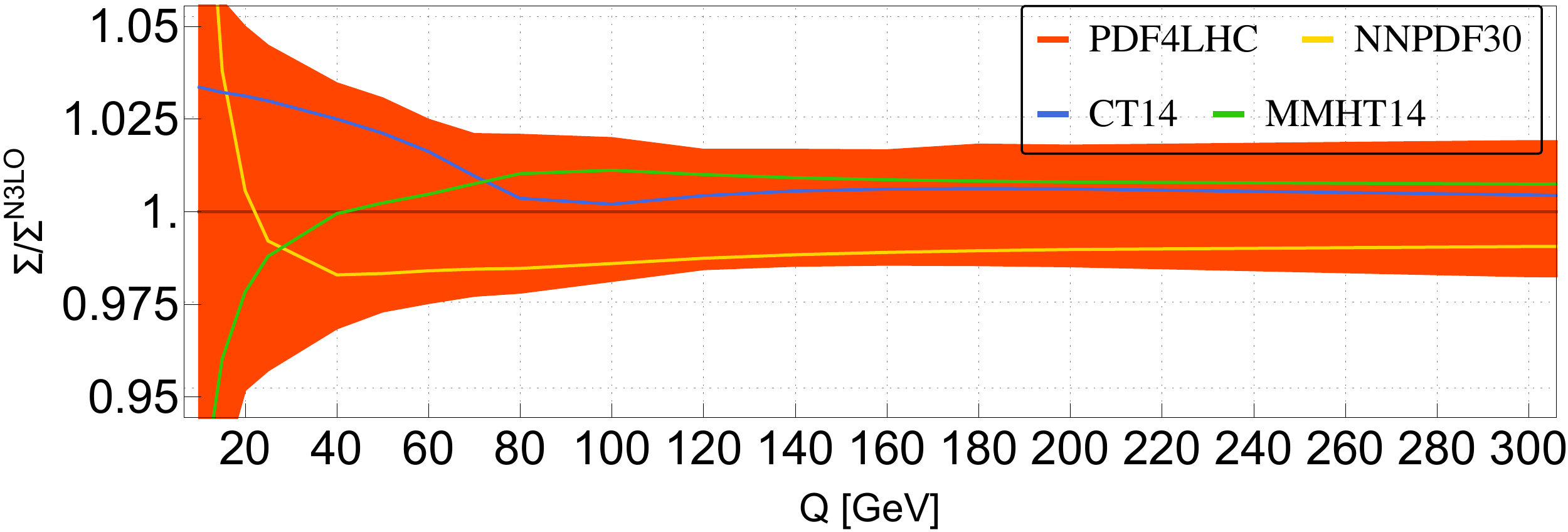}
\caption{\label{fig:old_PDFs}
Dependence of the NCDY process at N$^3$LO on the choice of the PDF set relative to the 
combined {\tt PDF4LHC15\_nnlo\_mc} set. The red band corresponds to the $\delta(\textrm{PDF})$ uncertainty. 
}
\end{figure*}
\begin{figure*}[!h]
\centering
\includegraphics[width=0.95\textwidth]{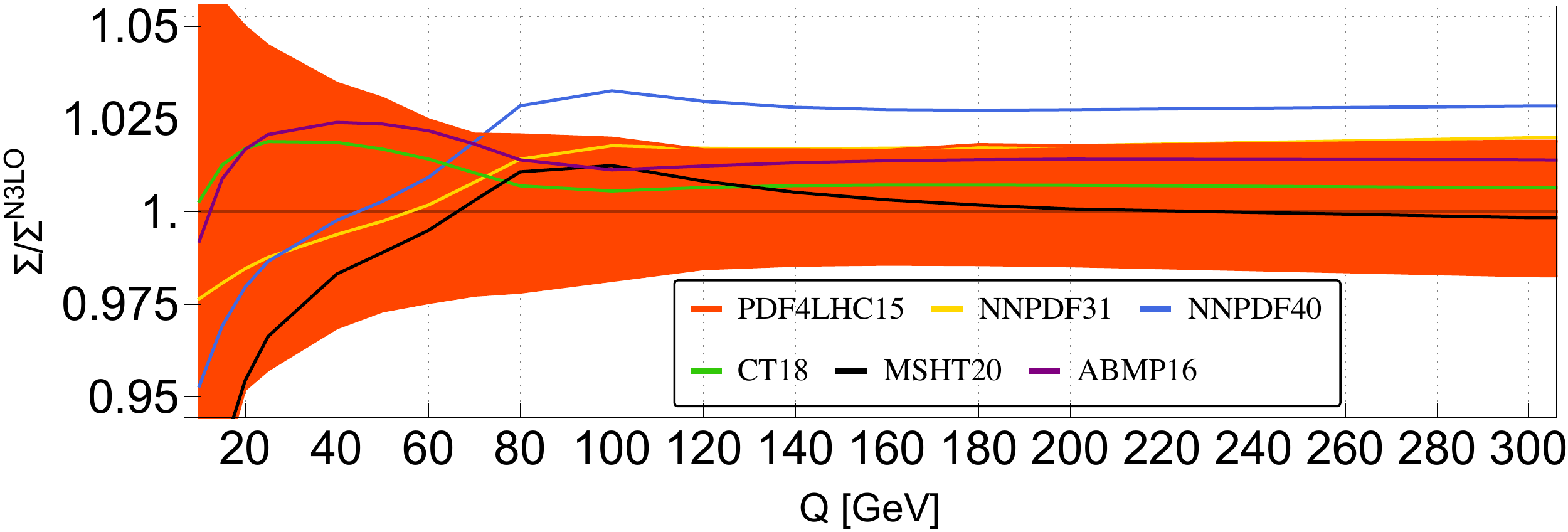}
\caption{\label{fig:new_PDFs}
Dependence of the NCDY process at N$^3$LO on the choice of the PDF set relative to the 
combined {\tt PDF4LHC15\_nnlo\_mc} set. The red band corresponds to the $\delta(\textrm{PDF})$ uncertainty. 
}
\end{figure*}
Figures~\ref{fig:old_PDFs} and~\ref{fig:new_PDFs} show the impact of evaluating the NCDY cross section with different PDF sets.
PDF4LHC15 is a combination of the CT14~\cite{Dulat:2015mca}, MMHT14~\cite{Harland-Lang:2014zoa} and NNPDF3.0~\cite{Ball:2014uwa} PDF sets and we show in fig.~\ref{fig:old_PDFs} predictions based on these individual sets relative to the prediction based on the {\tt PDF4LHC15\_nnlo\_mc} set.
The red band in fig.~\ref{fig:old_PDFs} reflects the $\delta(\text{PDF})$ uncertainty of PDF4LHC15 and we observe that the predictions based on the individual PDF sets are contained within this band and that their spread is comparable in size to this band.
Since the publication of the PDF4LHC15 combination a plethora of developments and the inclusion of LHC data into global PDF fits has led to updated PDF.
In fig.~\ref{fig:new_PDFs} we study in particular the PDF sets ABMP16~\cite{Alekhin:2016uxn}, CT18~\cite{Hou:2019efy}, MSHT20~\cite{Bailey:2020ooq}, NNPDF3.1~\cite{Ball:2017nwa} and NNPDF4.0~\cite{Ball:2021leu}.
We observe, that the spread among the newer sets is less than the the spread of their predecessors. 
In particular, in the range where $Q$ is comparable to the $Z$ and $W$-boson masses the different PDF sets seem to agree nicely. 
We observe furthermore that NNPDF4.0 leads to a significant enhancement of the NCDY cross section at large values of $Q$.
It would be interesting to study the impact of new PDF sets on high precision processes in further detail in the future and we are looking forward to an updated version of PDF4LHC~\cite{Cridge:2021qjj}.

\subsection{Contributions from interferences}
We conclude this phenomenological study by investigating some properties of the NCDY cross section at N$^3$LO and how it receives contributions from the photon, the $Z$-boson and their interference. 

\begin{figure*}[!h]
\centering
\includegraphics[width=0.95\textwidth]{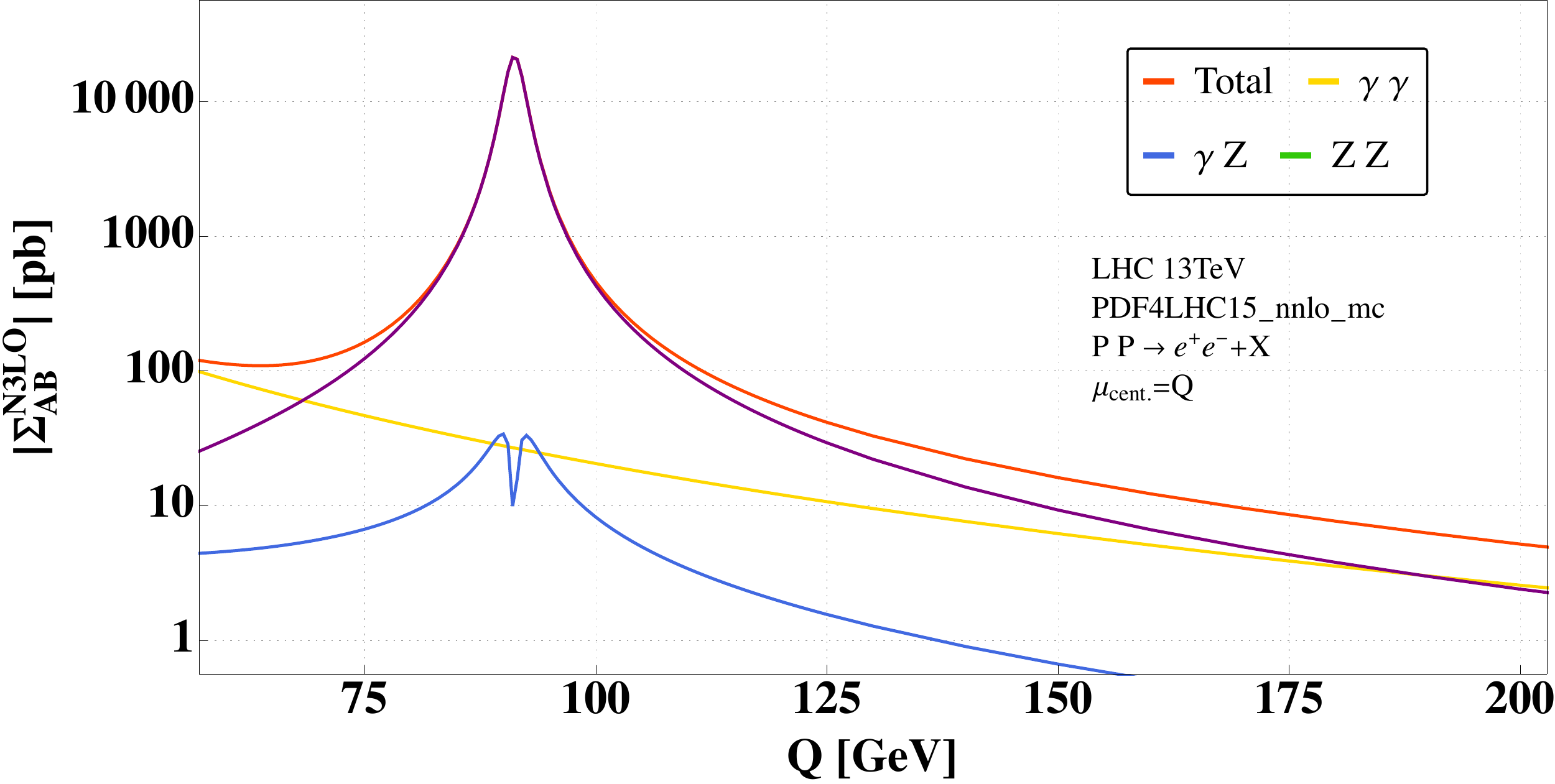}
\caption{\label{fig:composition_absolute}
Decomposition of the invariant-mass distribution into the absolute value of contributions from photon and $Z$ exchange and their interference as a function of $Q$ at different order in perturbations theory.
}
\end{figure*}
\begin{figure*}[!h]
\centering
\includegraphics[width=0.95\textwidth]{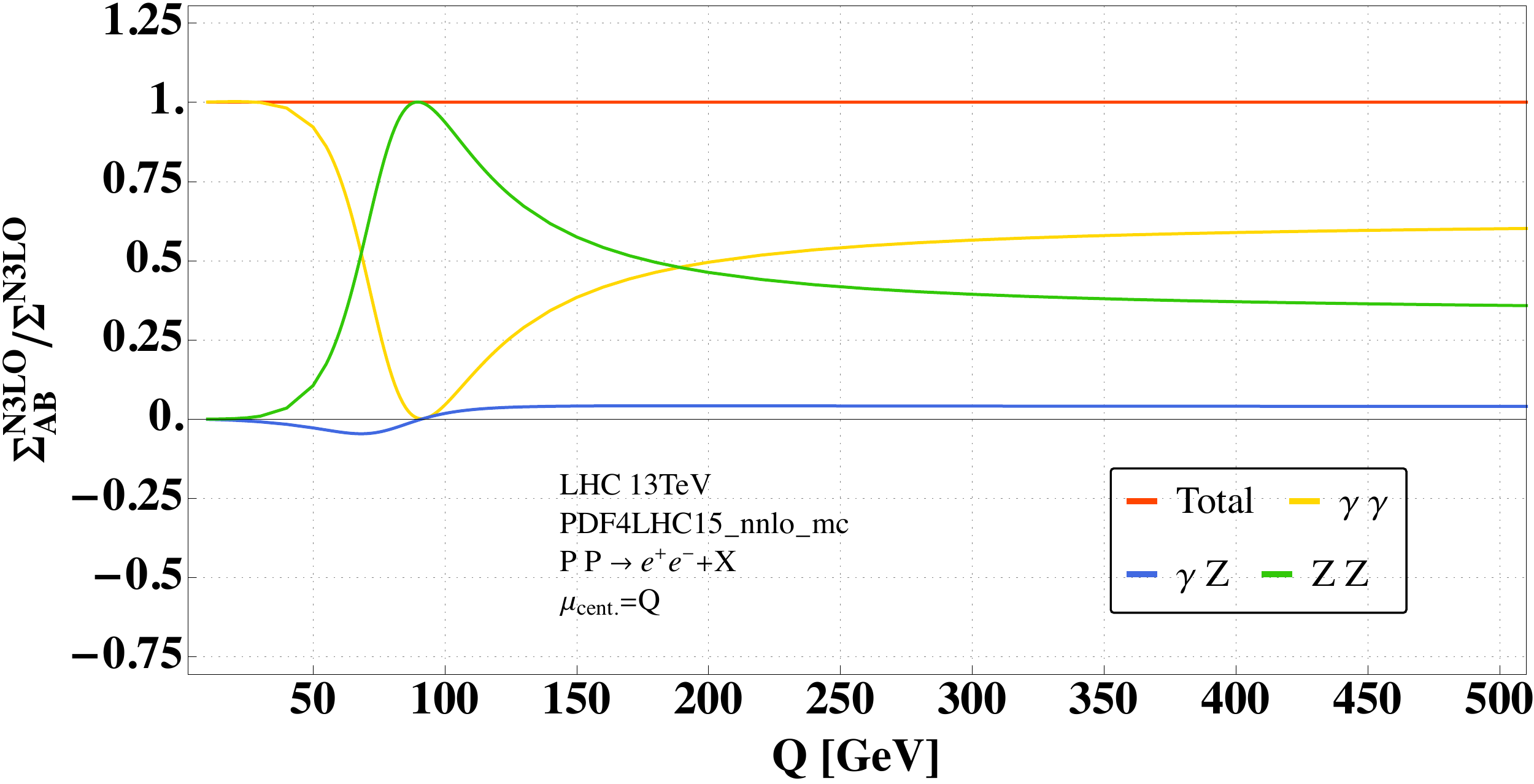}
\caption{\label{fig:composition_relative}
Relative contributions to the invariant-mass distribution from photon and $Z$ exchange and their interference as a function of $Q$ at different order in perturbations theory.
}
\end{figure*}
In figures~\ref{fig:composition_absolute} and~\ref{fig:composition_relative} we show the relative size of the three contributions. We see that, as expected, the photon-only contribution dominates below the $Z$-threshold. Above threshold all contributions are of similar size. Close to the threshold the cross section is dominated entirely by the $Z$ boson, and the interference changes sign when the threshold is crossed, as expected. 
The observed pattern is hardly modified by perturbative corrections.

Computations with $\gamma^5$ matrices are technically more involved than for purely vectorial couplings. 
For this reason, the axial-vector contribution is frequently approximated by simply using the same partonic coefficient functions as for the vector contribution and simply changing the coupling of the quarks in an appropriate fashion. 
While this is correct for contributions from Dirac traces with an even number of $\gamma^5$ matrices, a mismatch is introduced starting from NNLO for contributions involving traces with an odd number of $\gamma^5$ matrices. 
Moreover, effects from the axial-anomaly and the resulting non-decoupling of the top-quark are neglected in this way. 

\begin{figure*}[!h]
\centering
\includegraphics[width=0.9\textwidth]{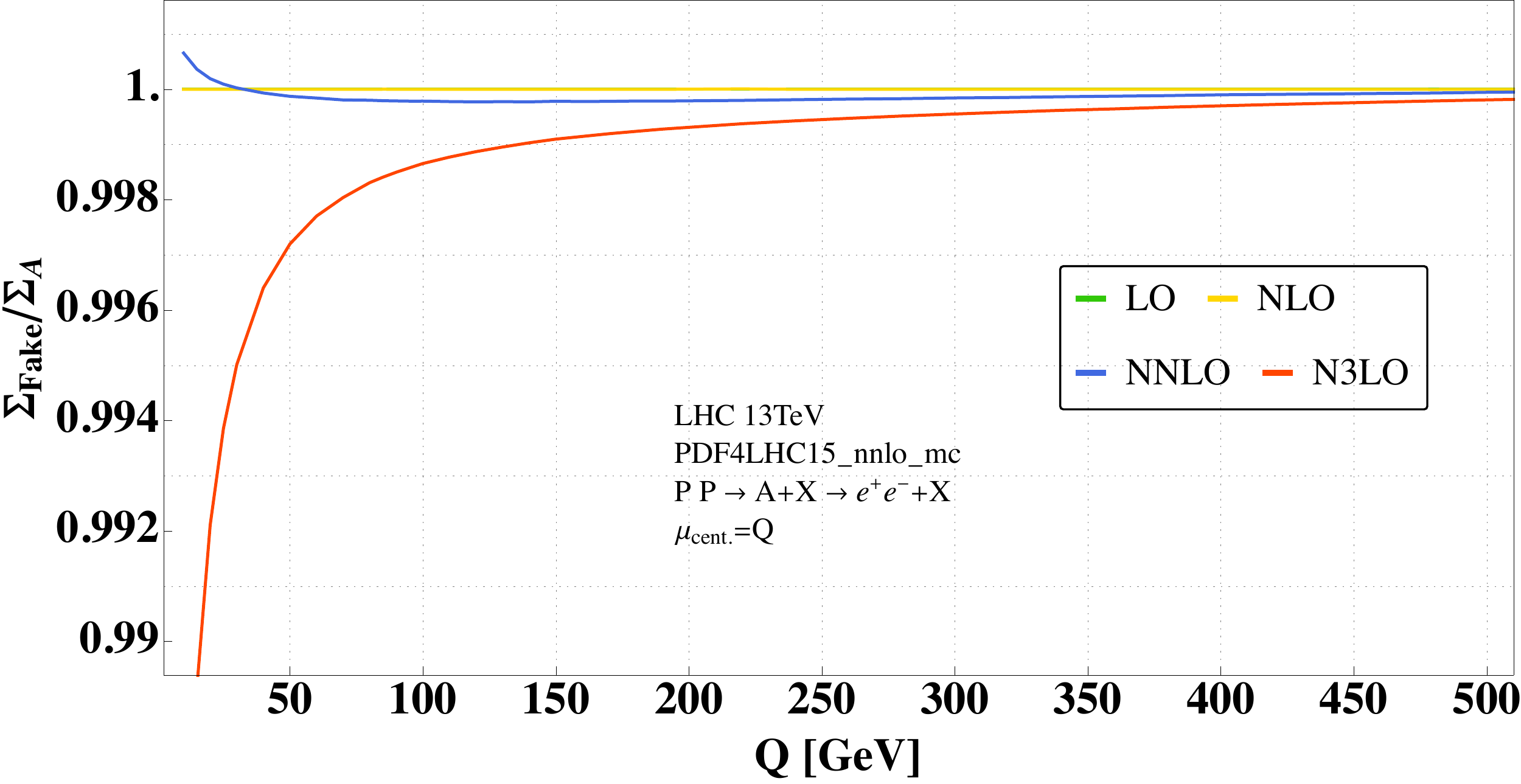}
\caption{\label{fig:all_fake}
Relative difference of approximating the axial-vector production cross section by a pure vector contribution as a function of $Q$ at different order in perturbations theory.
}
\end{figure*}
In figure~\ref{fig:all_fake} we study the impact of replacing the axial-vector coefficient functions $\eta^{A,i}_{\, ij\to A +X} $ in eq.~\eqref{eq:chargedecomp} by their vector counterparts $\eta^{V,i}_{\, ij\to Z +X} $ (while keeping the correct charges associated with the axial-vector). 
The figure displays the ratio of the contribution to the DY cross section due to the axial-vector current based on these `fake' partonic coefficient functions $\Sigma_{\text{Fake}}$ relativ to their correct contribution $\Sigma_{\text{A}}$.
We observe that at NNLO the mismatch introduced is negligible, and it is well below the permille level for the whole range of invariant masses considered. 
At N$^3$LO instead, the mismatch increases for small invariant masses, reaching the percent-level at invariant masses below 50 GeV. 
However, the numerical impact grows with decreasing $Q$ as the the relative contribution of the $Z$-boson diminishes and the photon exchange contribution dominates the DY production cross section, as can be seen from fig.~\ref{fig:composition_relative}.
We would like to point out that sufficiently differential DY-type cross sections are afflicted by differences of the vector and axial-vector contributions (see for example ref.~\cite{Dixon:1997th}) that integrate to zero in the inclusive cross section.

\begin{figure*}[!h!]
\centering
\includegraphics[width=0.9\textwidth]{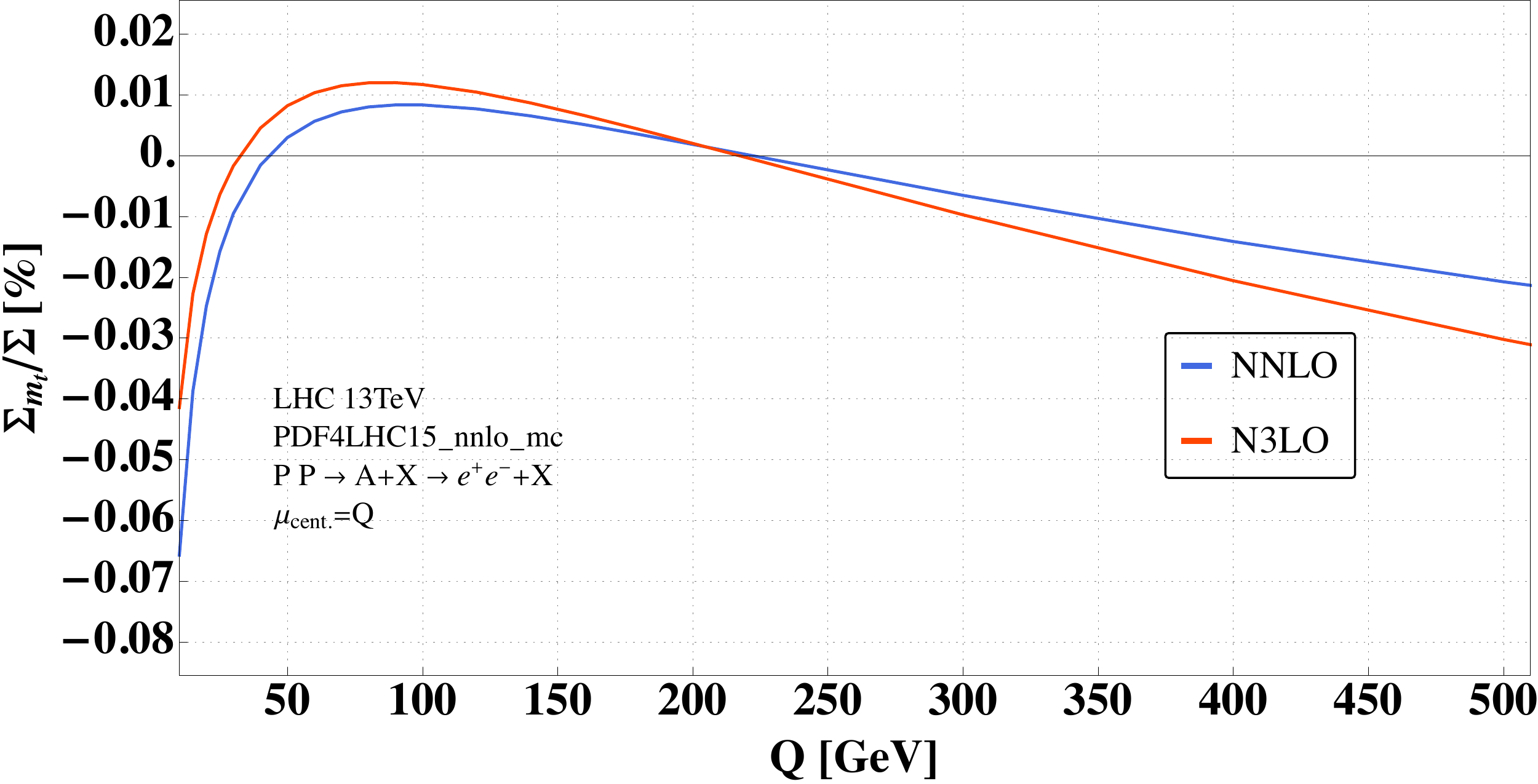}
\caption{\label{fig:top_WC}
Relative contribution from the non-decoupling top-mass effects as a function of $Q$ at different order in perturbations theory. 
}
\end{figure*}
In figure~\ref{fig:top_WC} we investigate the relative size of the non-decoupling of the top-quark, i.e., the effect of including or not the Wilson coefficient $C_{A,f}(\mu)$. We see that for the whole range of invariant masses considered, the effect of including the Wilson coefficient or not is well below the permille level.
\newpage


\section{Conclusions}
\label{sec:conclusions}

In this paper we have computed for the first time the complete inclusive N$^3$LO corrections in the strong coupling constant for the production of a massless lepton pair. This result has been made possible by combining our computation for the N$^3$LO corrections to vector production from ref.~\cite{Duhr:2020seh} with the computation of the axial-vector production cross section. A main ingredient in our computation is the use of the Larin scheme to treat the $\gamma^5$ matrix in dimensional regularisation. Our computation has passed several non-trivial checks: Apart from the explicit cancellation of all ultraviolet and infrared poles, we have checked that our implementation of the Larin scheme reproduces the contributions to the axial-vector cross section where both $\gamma^5$ matrices appear in the same Dirac trace, and that our final result satisfies the expected DGLAP renormalisation group equation once the non-decoupling top-mass effects are included. 
We include all partonic coefficient functions for the Drell-Yan production cross section as described by eq.~\eqref{eq:chargedecomp} as ancillary files together with the arXiv submission of this article.

We have studied the phenomenological impact of our computation. We find that the complete N$^3$LO corrections to the NCDY process, including the axial-vector contributions, has the same features as for the photon-only and $W$ cases of refs.~\cite{Duhr:2020seh,Duhr:2020sdp}. We find that the dependence of the complete NCDY process on the perturbative scales is similar to the production of a $\gamma^*$ or $W$. In particular, the bands obtained by varying the perturbative scales by a factor of two do not overlap when going from NNLO to N$^3$LO, showing once more the importance of considering N$^3$LO corrections for precision LHC observables. We have also considered ratios of K-factors between $\gamma^*$, $W$ and the complete NCDY process, and we find that, while the K-factors are very similar, the ratio depends on the invariant mass of the lepton pair, reaching a few percent depending on the invariant masses considered. This shows that care is needed when taking K-factors from one process and to apply it to another process, especially when aiming for precision results. Finally, we have also studied the impact of the choice of the PDF on our results. In particular, we have analysed how the missing N$^3$LO impact our predictions, which is one of the main bottlenecks to further improve theoretical predictions for LHC observables. Since the NCDY process is one of the main measurements used to constrain PDFs from LHC data, we expect that our computation will play an important role in the future to determine precisely the proton structure.

\section*{Acknowledgements}
We are grateful to W.-L. Ju, M. Sch\"onherr, L. Chen, M. Czakon and M. Niggetiedt for correspondence regarding the three-loop Wilson coefficient in refs.~\cite{Ju:2021lah,Chen:2021rft}.
We are grateful to Lance Dixon for useful discussions.
BM was supported by the Department of Energy, Contract DE-AC02-76SF00515.

\addcontentsline{toc}{section}{References}
\bibliographystyle{jhep}
\bibliography{refs}

\end{document}